\documentclass[prd,superscriptaddress,showpacs,nofootinbib,amsmath,amssymb]{revtex4}

\usepackage{bm}
\usepackage{amsfonts}
\usepackage{latexsym}
\usepackage[latin1]{inputenc}
\usepackage{graphicx}
\usepackage{amsmath}

\def \nn  {\nonumber}

\def\jnl@style{\it}
\def\aaref@jnl#1{{\jnl@style#1}}

\def\aaref@jnl#1{{\jnl@style#1}}

\def\aj{\aaref@jnl{AJ}}                   
\def\apj{\aaref@jnl{ApJ}}                 
\def\apjl{\aaref@jnl{ApJ}}                
\def\apjs{\aaref@jnl{ApJS}}               
\def\apss{\aaref@jnl{Ap\&SS}}             
\def\aap{\aaref@jnl{A\&A}}                
\def\aapr{\aaref@jnl{A\&A~Rev.}}          
\def\aaps{\aaref@jnl{A\&AS}}              
\def\mnras{\aaref@jnl{MNRAS}}             
\def\prd{\aaref@jnl{Phys.~Rev.~D}}        
\def\prl{\aaref@jnl{Phys.~Rev.~Lett.}}    
\def\qjras{\aaref@jnl{QJRAS}}             
\def\skytel{\aaref@jnl{S\&T}}             
\def\ssr{\aaref@jnl{Space~Sci.~Rev.}}     
\def\zap{\aaref@jnl{ZAp}}                 
\def\nat{\aaref@jnl{Nature}}              
\def\aplett{\aaref@jnl{Astrophys.~Lett.}} 
\def\apspr{\aaref@jnl{Astrophys.~Space~Phys.~Res.}} 
\def\physrep{\aaref@jnl{Phys.~Rep.}}      
\def\physscr{\aaref@jnl{Phys.~Scr}}       
\def\commat{\aaref@jnl{Comm.~Math.~Phys.}}		
\def\science{\aaref@jnl{Science}}		
\def\cqg{\aaref@jnl{Class.~Quantum Gravity}}		

\begin{document}

\title[Relativistic g-modes in rapidly rotating neutron stars]{Relativistic g-modes in rapidly rotating neutron stars}

\author{Erich Gaertig}
\affiliation{Theoretical Astrophysics, Eberhard-Karls University of T\"ubingen, T\"ubingen 72076, Germany}
\author{Kostas D. Kokkotas} 
\affiliation{Theoretical Astrophysics, Eberhard-Karls University of T\"ubingen, T\"ubingen 72076, Germany}
\affiliation{Department of Physics, Aristotle University of Thessaloniki, Thessaloniki 54124, Greece}
\date{\today}

\begin{abstract}
We study the g-modes of fast rotating stratified neutron stars in the general relativistic Cowling approximation, where we neglect metric perturbations and where the background models take into account the buoyant force due to composition gradients. This is the first paper studying this problem in a general relativistic framework.  In a recent paper \cite{Passamonti:2009zr}, a similar study was performed within the Newtonian framework, where the authors presented results about the onset of CFS-unstable g-modes and the close connection between inertial- and gravity-modes for sufficiently high rotation rates and small composition gradients. This correlation arises from the interplay between the buoyant force which is the restoring force for g-modes and the Coriolis force which is responsible for the existence of inertial modes. In our relativistic treatment of the problem, we find an excellent qualitatively agreement with respect to the Newtonian results.
\end{abstract}


\maketitle

\section{Introduction}
\label{sec:introduction}

Stellar oscillations serve as a unique tool in revealing the nature of the stellar interior and there is a collaborative effort during the last three decades  to collect this type of information from astrophysically important stars \citep{1996Sci...272.1284H, 2008JPhCS.118a2041G}. Actually, the Sun's seismic activity has been studied in more detail due to its proximity and thus there already exists an independent branch in astrophysics called helioseismology. Since the mid 90s, it has been shown that asteroseismology can be combined with gravitational waves in order to shed light in the interior structure  of relativistic stars \citep{Andersson:1996pn,Andersson:1997rn,Kokkotas:1999mn,Benhar:2004xg}. This initial work has been extended by many groups and even the most exotic cases of compact objects like quark stars \citep{Sotani:2003bg}  or compact stars in alternative theories of gravity \citep{Sotani:2005qx,Sotani:2009xw} have been studied. In addition, more detailed microphysics has been taken into account, for example, stars with superfluid components \citep{Andersson:2006nr, Andersson:2008sf} or even crust and magnetic fields \citep{Sotani:2006at,Samuelsson:2006tt}.

Rotation enriches the stellar oscillation spectrum with new frequencies and many complications. This is mainly due to the breaking of the spherical symmetry of the problem which complicates the form of the oscillation equations and adds to the oscillation spectrum with new families of oscillation modes, such as the so-called inertial modes, while it influences significantly all other families. In the general relativistic framework, due to the emission of gravitational radiation, there appears a new type of instability which is present only in rotating stars, the so-called CFS instability \citep{Chandrasekhar:1970rt,Friedman:1978fr}, for a review see \cite{2001IJMPD..10..381A}. Thus the rotational modes, r-modes, are unstable to any rotational rate while other types of modes like the f- and the g-modes are becoming  unstable only for fast rotating neutron stars. Especially, the so-called g-modes of rotating stars are present only in stars with internal stratification which may be associated with composition variations \citep{1992ApJ...395..240R} and are expected to be important for nascent neutron stars. The g-modes have been studied in detail mainly for non-rotating neutron stars in the Newtonian and the  general relativistic framework 
\citep{1986MNRAS.222..393F,1987MNRAS.227..265F, 2003MNRAS.338..389M,2007CQGra..24.5093F}.  

As it has been mentioned earlier, the effect of rotation on the perturbation equations is dramatic and the majority of studies on stellar oscillation and instabilities has been done using either the so-called  `\,slow rotation approximation\,' or using Newtonian theory \citep{2001IJMPD..10..381A,2003LRR.....6....3S}. The slow rotation approximation assumes that the star is rotating at such a rate that the effects of rotation on the shape of the star can be neglected. Only in the last five years or so it became possible to study fast rotating stars in the general relativistic framework either using a linearized form of relativistic perturbation equations \citep{2005MNRAS.356..217Y,Boutloukos:2006cx,Gaertig:2008kx} or by evolving the non-linear hydrodynamical equations in the relativistic framework
\citep{Stergioulas:2004fr,Dimmelmeier:2006zr,Kastaun:2006kx}. 

The most recent and detailed study of effects of rotation on g-modes by \cite{Passamonti:2009zr} assumes Newtonian gravity but it deals properly with the rotation, i.e. the stellar models are `\,fast\,' rotating. In particular, it has been demonstrated in this paper, that the g-modes for high rotation rates become rotationally dominated by the Coriolis force and they approach particular inertial modes. In our work, we perform a similar study but in the general relativistic framework  which describes better both the equilibrium and the perturbed neutron star configurations. Still, one does not expect dramatic qualitative changes as one moves from Newton's theory to general relativity, especially for the study of the g- and inertial modes and this is what we demonstrate here. Moreover, the Cowling approximation that has been used here is known already from the '60s, thanks to the pioneering work of Robe \cite{Robe:1968mz}, to be an excellent approximation for the study of g-modes and the higher p-modes but it is only qualitatively correct for the study of the lower p-modes and the f-mode.

In the next two sections we present the equations that describe the problem together with the boundary conditions of the problem, while in Section 4 and 5 we review the numerical techniques that have been used and describe the results of our simulations. Finally, in the last section we summarize the conclusions and suggest further work. 

\section{Problem setup}
\label{sec:setup}
The basic formulation of this work was already drafted in a previous article by the authors, see \cite{Gaertig:2008kx} (abbreviated GK08 in the remainder of this paper) for the case of barotropic oscillations. Here, we will shortly summarize the fundamental points and the extensions that were made in order to handle stratified stars.

In the relativistic Cowling approximation, where one neglects the perturbations of the spacetime, the hydrodynamic equations that govern linear oscillations around a static equilibrium take the form
\begin{equation}
\label{eq:perturbedEinstein}
\nabla^{\mu}(\delta T_{\mu\nu}) = g^{\mu\kappa}(\partial_{\mu}\delta T_{\kappa\nu}-{\Gamma^{\lambda}}_{\kappa\mu}\delta T_{\lambda\nu}
-{\Gamma^{\lambda}}_{\mu\nu}\delta T_{\kappa\lambda}) = 0\,,
\end{equation}
where $g^{\mu\nu}$ is the contravariant metric tensor of the background model, $\delta T_{\mu\nu}$ is the perturbed energy-momentum--tensor and ${\Gamma^{\lambda}}_{\kappa\mu}$ are the Christoffel symbols. We further make the assumption that the matter has no viscosity or shear stresses. In this case it can be described by a perfect fluid and $\delta T_{\mu\nu}$ has the form
\begin{equation}
\label{eq:deltaTmunu}
\delta T_{\mu\nu}=(\epsilon+p)(u_{\mu}\delta u_{\nu}+u_{\nu}\delta u_{\mu})+(\delta p+\delta\epsilon)u_{\mu}u_{\nu}+\delta p g_{\mu\nu}\,.
\end{equation}
Here $\epsilon$ is the energy-density, $p$ is the pressure, $u_{\mu}$ is the 4-velocity and $\delta p$, $\delta\epsilon$, $\delta u_{\mu}$ are its corresponding perturbations. Energy density and pressure are not independent from each other but are related via an equation of state (EoS) which we assume to be polytropic, i.e.
\begin{equation}
\label{eq:polytropicEoS}
p = K\rho^{1 + 1/N}\quad \mbox{where} \quad
\epsilon = \rho +Np\,.
\end{equation}
Here $\rho$ is the rest-mass density, $K$ the polytropic constant, $N$ the polytropic exponent and $\Gamma = 1 + 1/N$ the polytropic index.

The system of equations \eqref{eq:perturbedEinstein} leads to four relations between the six unknown quantities $\delta u_{\mu}$, $\delta\epsilon$ and $\delta p$. The normalization constraint $u^{\mu}\delta u_{\mu} = 0$ gives an additional relationship (which takes a very simple form especially in a comoving coordinate system, that is $\delta u_t = 0$) but still the system is underdetermined. What is needed is an auxiliary correlation between the Eulerian pressure perturbation $\delta p$ and the energy density perturbation $\delta\epsilon$. For adiabatic oscillations and a polytropic equation of state for both the background and the perturbation this relation is given by (see \cite{Ruoff:2003yq} for example)
\begin{equation}
\label{eq:fluidDisplacement}
\delta p = \frac{\Gamma_{1} p}{\epsilon + p}\delta\epsilon + p_{,\varrho}\xi^{\varrho}\left(\frac{\Gamma_{1}}{\Gamma} -1\right) + p_{,\zeta}\xi^{\zeta}\left(\frac{\Gamma_{1}}{\Gamma} -1\right)\,.
\end{equation}
Here $\Gamma_{1}$ is the polytropic index of the perturbed fluid elements and $\xi^{\varrho}$, $\xi^{\zeta}$ are the two relevant components of the fluid displacement vector $\xi^{\mu}$ in cylindrical coordinates. Since the stationary background is axisymmetric, there is no pressure variation with respect to $\varphi$ and hence no additional term proportional to $\xi^{\varphi}$ in equation \eqref{eq:fluidDisplacement}. In the barotropic case when $\Gamma = \Gamma_{1}$, the square of the speed of sound is given by $c_{s}^{2} = \Gamma p/(\epsilon + p)$ and relation \eqref{eq:fluidDisplacement} reduces to the well known form $\delta p = c_{s}^2\delta\epsilon$ also used in GK08.

We now have a connection between $\delta p$ and $\delta\epsilon$ but at the cost of introducing two new quantities, i.e. the fluid displacement components, that have to be evolved in time as well. There has to be a relation between $\delta u_{\mu}$ and fluid displacement $\xi^{\mu}$ since the latter is basically an integrated velocity. In the Cowling approximation this dependency is given by (see e.g. \cite{Ruoff:2002vn})
\begin{equation}
\label{eq:evolEqForFluidDisplacement}
\delta u_{\mu} = (g_{\mu\nu} + u_{\mu}u_{\nu})L_{u}\xi^{\nu}\,,
\end{equation}
where $L_{u}$ denotes the Lie derivative along $u^{\mu}$. This Lie derivative will introduce a time derivative on the righthand side of equation \eqref{eq:evolEqForFluidDisplacement} which then will serve as an evolution equation for the components of the fluid displacement vector $\xi^{\mu}$. 

\section{Evolution Equations and Boundary Conditions}
\label{sec:evolutionEquations}
As in GK08 we are using a comoving frame of reference in cylindrical coordinates $(\varrho, \zeta, \varphi,t)$. In such a system the metric reads
\begin{equation}
\label{eq:lineElement}
ds^{2} = e^{-2U}\left[e^{2k}\left(d\varrho^{2}+d\zeta^{2}\right)+W^{2}d\varphi^{2}\right]-e^{2U}(dt+a d\varphi)^{2}\,,
\end{equation}
where $U$, $k$, $W$ and $a$ are axisymmetric metric potentials that depend on $\varrho$ and $\zeta$ only. The $(\varphi,t)$-component of the metric tensor is proportional to the potential $a$ and vanishes in the absence of rotation.

With the polytropic equation of state in the form of relation \eqref{eq:polytropicEoS}, the polytropic index $\Gamma$ is equivalent to the adiabatic index and it is
\begin{equation*}
\Gamma = \frac{\epsilon + p}{p}\frac{dp}{d\epsilon}\,.
\end{equation*}
On the other hand, for stratified systems the speed of sound $c_{s}$ is defined by
\begin{equation*}
c_{s}^{2} = \frac{\Gamma_{1}}{\Gamma}\frac{dp}{d\epsilon}\,.
\end{equation*}
Using the last two equations one finally arrives at
\begin{equation*}
c_{s}^{2} = \frac{\Gamma_{1}p}{\epsilon + p}
\end{equation*}
for stratified stars. This allows us to rewrite equation \eqref{eq:fluidDisplacement} into
\begin{equation}
\label{eq:pressurePerturbation}
\delta p = c_{s}^{2}\delta\epsilon + c_{s}^{2}\tilde{g}_{1} + c_{s}^{2}\tilde{g}_{2}\,,
\end{equation}
where we definded
\begin{equation}
\label{eq:defOfG}
\tilde{g}_{1} = \frac{p_{,\varrho}\xi^{\varrho}}{c_{s}^{2}}\left(\frac{\Gamma_{1}}{\Gamma} -1\right)\quad \mbox{and} \quad \tilde{g}_{2} = \frac{p_{,\zeta}\xi^{\zeta}}{c_{s}^{2}}\left(\frac{\Gamma_{1}}{\Gamma} -1\right)\,.
\end{equation}
Since both relations \eqref{eq:perturbedEinstein}, \eqref{eq:pressurePerturbation} are linear and equation \eqref{eq:pressurePerturbation} reduces to the barotropic case for $\tilde{g}_{1} = \tilde{g}_{2} = 0$, we expect to see additional terms in the resulting set of time-evolution equations which are similar to the corresponding expressions for the energy density perturbation $\delta\epsilon$ already obtained in the barotropic version of the problem. This will also serve as a simple check of our calculations.

For fast rotating configurations, it is not reasonable any more to decompose the fluid perturbations into spherical harmonics. Centrifugal forces, though only an $\mathcal{O}(\Omega^2)$-effect where $\Omega$ is the rotation rate of the star,  will lead to a flattening of the rapidly spinning object where spherical harmonics are no longer appropriate as a system of base functions. Instead, we separate the azimuthal dependence into complex exponential functions and we write
\begin{eqnarray}
\label{eq:perturbedAnsatzFluid}
(\epsilon + p)W e^{U}\,\delta u_{\varrho}&=&f_{1}(\varrho,\zeta,t)e^{im\varphi}\nn\\
(\epsilon + p)W e^{U}\,\delta u_{\zeta}&=&f_{2}(\varrho,\zeta,t)e^{im\varphi}\\
(\epsilon + p)\,\delta u_{\varphi}&=&f_{3}(\varrho,\zeta,t)e^{im\varphi}\nn\\
c_{s}^{2}e^{U}\,\delta\epsilon&=&H(\varrho,\zeta,t)e^{im\varphi}\nn
\end{eqnarray}
for the standard fluid perturbation variables and
\begin{eqnarray}
\label{eq:perturbedAnsatzDisplacement}
c_s^2 e^{U}\,\tilde{g}_1 &=& g_{1}(\varrho,\zeta,t)e^{im\varphi}\\
c_s^2 e^{U}\,\tilde{g}_2 &=& g_{2}(\varrho,\zeta,t)e^{im\varphi}\nn
\end{eqnarray}
for the displacement vector variables. Inserting \eqref{eq:pressurePerturbation} into \eqref{eq:deltaTmunu}, using equations \eqref{eq:perturbedAnsatzFluid}, \eqref{eq:perturbedAnsatzDisplacement} and the normalization constraint $\delta u_{t} = 0$ for replacing $\delta T_{\mu\nu}$ in relation \eqref{eq:perturbedEinstein}, finally leads to the system of evolution equations for the fluid perturbations in stratified stars:
\begin{eqnarray}
\label{eq:explicit_v2}
\frac{\partial{f}_{1}}{\partial t}& = &-We^{U}\left(\frac{\partial{H}}{\partial \varrho} +\frac{\partial{g_{1}}}{\partial \varrho} + \frac{\partial{g_{2}}}{\partial \varrho}\right)
-\frac{e^{5U}}{W}\frac{\partial a}{\partial \varrho}{f}_{3}
-\frac{W}{c_{s}^{2}}\frac{\partial U}{\partial \varrho}e^{U}{H}\nn\\
\frac{\partial{f}_{2}}{\partial t}& = &-We^{U}\left(\frac{\partial{H}}{\partial\zeta} + \frac{\partial{g_{1}}}{\partial\zeta} + \frac{\partial{g_{2}}}{\partial\zeta}\right)
-\frac{e^{5U}}{W}\frac{\partial a}{\partial \zeta}{f}_{3}
-\frac{W}{c_{s}^{2}}\frac{\partial U}{\partial \zeta}e^{U}{H}\nn\\
\frac{\partial{f}_{3}}{\partial t}& = &\frac{im}{F}\left(ac_{s}^{2}e^{4U}{f}_{3}+W^{2}(H+g_{1}+g_{2})\right) +\frac{Wac_{s}^{2}e^{3U-2k}}{F}\left(\frac{\partial{f}_{1}}{\partial \varrho}+\frac{\partial{f}_{2}}{\partial \zeta}\right) -\frac{e^{3U-2k}}{F}W\frac{\partial a}{\partial \varrho}{f}_{1}
-\frac{e^{3U-2k}}{F}W\frac{\partial a}{\partial \zeta}{f}_{2}\\
& & -\frac{A a W  e^{3U-2k}}{(\epsilon + p)F} f_{1} -\frac{B a W  e^{3U-2k}}{(\epsilon + p)F} f_{2}\nn\\
\frac{\partial{H}}{\partial t}& = &\frac{im}{F}\left(c_{s}^{2}e^{4U}{f}_{3} + ac_{s}^{2}e^{4U}(H + g_{1} + g_{2})\right) +\frac{Wc_{s}^{2}e^{3U-2k}}{F}\left(\frac{\partial{f}_{1}}{\partial \varrho}+\frac{\partial{f}_{2}}{\partial \zeta}\right) -\frac{c_{s}^{2}e^{7U-2k}}{F}\frac{a}{W}\frac{\partial a}{\partial \varrho}{f}_{1}-\frac{c_{s}^{2}e^{7U-2k}}{F}\frac{a}{W}\frac{\partial a}{\partial \zeta}{f}_{2}\nn\\
& & -\frac{A a^{2}c_{s}^{2}e^{7U-2k}}{(\epsilon + p)F W}f_{1} -\frac{B a^{2}c_{s}^{2}e^{7U-2k}}{(\epsilon + p)F W}f_{2}\,,\nn
\end{eqnarray}
where
\begin{equation*}
\label{eq:Fdefinition}
F:=a^{2}c_{s}^{2}e^{4U}-W^{2}
\end{equation*}
is just a convenient abbreviation whereas the coeffients
\begin{equation*}
A:=p_{, \varrho}\left(\frac{\Gamma_1}{\Gamma} - 1\right)\quad , \quad B:=p_{, \zeta}\left(\frac{\Gamma_1}{\Gamma} - 1\right)
\end{equation*}
are closely related to the vector field
\begin{equation}
\label{eq:SchwarzschildDiscriminant}
\mathbf{A} = \left(\frac{1}{\Gamma} - \frac{1}{\Gamma_1}\right)\nabla\ln p
\end{equation}
whose amplitude is referred to as Schwarzschild discriminant and which determines the convective stability of g-modes. For $|\mathbf{A}| = 0$ we are again in the barotropic limit while $|\mathbf{A}| < 0$ means convective stability and $|\mathbf{A}| > 0$ signals unstable g-modes (see e.g. \cite{Unno:1989kx}).

A couple of observations can be made here. Firstly, if we restrict the problem to barotropic perturbations again, then $g_{1} = g_{2} = A = B = 0$ and the evolution equations \eqref{eq:explicit_v2} reduce to the system of equations successfully implemented already in GK08. Secondly, the appearance of terms like $H + g_1 + g_2$ in the last two equations of system \eqref{eq:explicit_v2} as well as its derivatives with respect to $\varrho$ and $\zeta$ in the first two equations is a direct consequence of the linearity of the perturbation equations and the decomposition made in equation \eqref{eq:pressurePerturbation}.

The system \eqref{eq:explicit_v2} is still incomplete though. On the righthand side, the fluid displacement varibles $g_1$ and $g_2$ appear and one needs evolution equations for them as well in order to update these quantities for the next time step. This is done with the help of relation \eqref{eq:evolEqForFluidDisplacement}. Since the Lie derivative of a function agrees with the standard differentiation of a scalar variable along a vector field and since $u^{\mu}$ has a very simple form in comoving cylindrical coordinates, i.e. $u^{\mu} = \left(0,0,0,e^{-U}\right)$, one immediately arrives at the evolution equations for the fluid displacement quantities which take the form
\begin{eqnarray}
\label{eq:eqForDisplacement}
\frac{\partial g_1}{\partial t} & = & \frac{A e^{3U -2k}}{(\epsilon + p)W}f_{1}\\
\frac{\partial g_2}{\partial t} & = & \frac{B e^{3U -2k}}{(\epsilon + p)W}f_{2}\nn\,.
\end{eqnarray}
As already discussed in Section \ref{sec:setup}, these equations basically describe that the fluid displacement is obtained by integrating the corresponding velocities. Actually, relationship \eqref{eq:eqForDisplacement} has already been used for deriving the evolution equations for $f_3$ and $H$, since on their righthand side additional time derivatives of $g_1$ and $g_2$ appeared; i.e. the coefficients proportional to $A$, $B$.

The system of evolution equations is closed by the appropriate boundary conditions that prescribe the behaviour of the perturbations at the  boundaries of the numerical domain. Using the same computational setup as described in GK08, these boundaries consist of the rotation axis and the surface of the star. The original boundary conditions for the barotropic evolution variables do not change at all in the case of stratification and a very similar analysis can be carried out for the additional quantities $g_1$, $g_2$. Let us first consider the surface of the star. Restricting ourselves to linear perturbations, both variables $\tilde{g}_1$ and $\tilde{g}_2$ in equation \eqref{eq:pressurePerturbation} have to be finite in order to add up to a finite Eulerian pressure perturbation $\delta p$ at the surface. By virtue of equation \eqref{eq:perturbedAnsatzDisplacement} this means, that $g_1$ and $g_2$ have to vanish there. The boundary condition for all evolution quantities along the surface therefore is
\begin{eqnarray}
\label{eq:bcSurface}
f_{1|\mathrm{surface}} = f_{2|\mathrm{surface}} = f_{3|\mathrm{surface}} = H_{|\mathrm{surface}} &=& 0\\
g_{1|\mathrm{surface}} = g_{2|\mathrm{surface}} &=& 0\,.\nn
\end{eqnarray}
For the rotation axis, one has to distinguish between scalar and vectorial perturbations the same way it was done in the barotropic case. Scalar perturbations have to be unique along the axis for all values of $m$ when varying the azimuthal angle $\varphi$ while vectorial perturbations are only allowed to change like $\cos(\varphi)$ or $\sin(\varphi)$ near the axis. Put in other words, only for $m = \pm 1$ they can attain nonzero values there. In the case presented here, $g_1$ which represents $\xi^{\varrho}$ clearly has vectorial character while $g_2$ can be treated as a scalar quantity concerning the proper behaviour along the rotation axis. Table \ref{tab:bcAxis} gives a complete overview of the boundary conditions for all evolution quantities.

\begin{table}
\centering
\caption{Boundary conditions for the perturbation quantities along the rotation axis. Here, 'f\&c' stands for 'finite and continuous'.}
\label{tab:bcAxis}
\begin{tabular}{ccccccc}
\hline
$m$-value & $f_{1|\mathrm{axis}}$ & $f_{2|\mathrm{axis}}$ & $f_{3|\mathrm{axis}}$ & $H_{|\mathrm{axis}}$ & $g_{1|\mathrm{axis}}$ & $g_{2|\mathrm{axis}}$\\
\hline
0 & 0 & 0 & 0 & f\&c & 0 & f\&c\\
$\pm 1$ & 0 & 0 & f\&c & 0 & f\&c & 0\\
else & 0 & 0 & 0 & 0 & 0 & 0\\
\hline
\end{tabular}
\end{table}

\section{Numerical Implementation}
\label{sec:numericalImplementation}
The implementation to solve the system of equations \eqref{eq:explicit_v2} is virtually  adopted from GK08. Just like in that paper, a pseudospectral code is used to generate the axisymmetric background model which then gets interpolated onto a finite difference grid that extends the original computational domain in polar direction to $0 \leq\theta\leq\pi$. The time evolution equations themselves are solved by an Iterated Crank-Nicholson scheme with swapped weights (\cite{Teukolsky:2000rt,Leiler:2006ys}) and a weighting factor of $\alpha = 0.6$. Similar to the barotropic case, the code is numerically unstable. Exponentially growing modes destroy the stability already after very few timesteps. A dissipative, second order Kreiss-Oliger term was added to the fluid displacement variables $g_1$ and $g_2$ as well. However, it was found that the amount of dissipation needed for $g_1$, $g_2$ is at least two orders of magnitude smaller than for the rest of the evolution variables. With this artificial viscosity, the numerical evolution stays stable for all simulations shown in this paper. The typical evolution time was around 150 ms; this leads to a frequency resolution of $\Delta f\approx 7$ Hz.

In addition, mode recycling routines as in \cite{Stergioulas:2004fr}, \cite{Kastaun:2006kx} and GK08 were used to identify and enhance specific modes as well as for the extraction of eigenfunctions.

\section{Results}
\label{sec:results}
\subsection{Background Models and Initial Data}
\label{ssec:backgroundModels}

In this paper, we use the polytropic BU model series (\cite{Font:2001kx}) for which $K = 100$, $\Gamma = 2$ and a fixed central rest-mass density of $\rho_{c} = 1.28\times 10^{-3}$ in units of $G = c = M_{\odot} = 1$. In the convential cgs-system this means a central rest-mass density of $\rho_{c} = 0.79\times 10^{15}$ g/cm$^3$ and a corresponding energy density of $\epsilon_{c} = 0.89\times 10^{15}$ g/cm$^3$. In the nonrotating limit this leads to a neutron star with `\,standard values\,' for gravitational mass, $M = 1.4\,M_{\odot}$, and circumferential radius, $r = 14.16$ km, and is labelled BU0. Starting from there, the angular velocity is subsequently increased in such a way that  the ratio of polar to equatorial radius $r_p/r_e$ decreases by a factor of $0.05$ for every new equilibrium configuration until the mass-shedding limit of $\Omega = 5.36$ kHz is reached. Table \ref{tab:buModels} gives an overview over the model parameters for this specific equation of state.

In order to resolve the splitting of the co- and counterrotating g-mode oscillations accurately at low frequencies, additional equilibrium configurations were taken into account between the nonrotating model and BU1. Since the latter one already has a spin frequency of around $40\%$ of the Kepler-limit, three auxiliary equilibrium models were constructed to trace the regime of low rotation rates.

\begin{table}
\centering
\caption{Background model parameters; $\Omega$ is the rotation rate, $M$ the gravitational mass and $r_p/r_e$ the ratio of polar to equatorial circumferential radius}
\label{tab:buModels}
\begin{tabular}{lcccc}
\hline
model & $\Omega$ (kHz) & $r_p/r_e$ & M ($M_\odot$) & $r_e$ (km)\\
\hline
BU0 		& 0.0		& 1.0		& 1.4		& 14.16 \\
BU01	& 0.694 	& 0.995	& 1.4		& 14.19	\\ 
BU02	& 0.981	& 0.99	& 1.41	& 14.22	\\
BU03	& 1.695	& 0.97	& 1.42	& 14.36	\\
BU1   	& 2.182	& 0.95	& 1.43	& 14.52\\
BU2   	& 3.062	& 0.9	& 1.47	& 14.92\\
BU3   	& 3.712	& 0.85	& 1.5	& 15.39\\
BU4   	& 4.229	& 0.8	& 1.54	& 15.91\\
BU5   	& 4.647	& 0.75	& 1.59	& 16.53\\
BU6   	& 4.976 	& 0.7	& 1.63	& 17.27\\
BU7   	& 5.214	& 0.65	& 1.67	& 18.16\\
BU8   	& 5.344	& 0.6	& 1.69	& 19.29\\
\hline
\end{tabular}
\end{table}

Concerning initial data, we are following \cite{Passamonti:2009zr} and mostly use a Gaussian radial profile in the pressure perturbation of the form
\begin{equation}
\label{eq:initialData}
\delta p = A\rho\,\mathrm{exp}{\left(\frac{r - r_0}{qr_s(\theta)}\right)^2}Y_{lm}(\theta,\phi)\,.
\end{equation}
Here, $A$ is a dimensionless amplitude factor, $\rho$ the rest-mass density, $r_s(\theta)$ the coordinate radius of the star at polar angle $\theta$ and $r_0$, $q$ determine the centre and width of the Gaussian perturbation. This type of initial data typically excites not only g-modes but also pressure- and inertial modes. Applying mode-recycling techniques is crucial to extract, identify and follow specific modes throughout the range of possible spin frequencies.

\subsection{Barotropic Oscillations}
\label{ssec:barotropicOscillations}
\begin{figure}[ht!]
\centering
\includegraphics[width=0.49\textwidth]{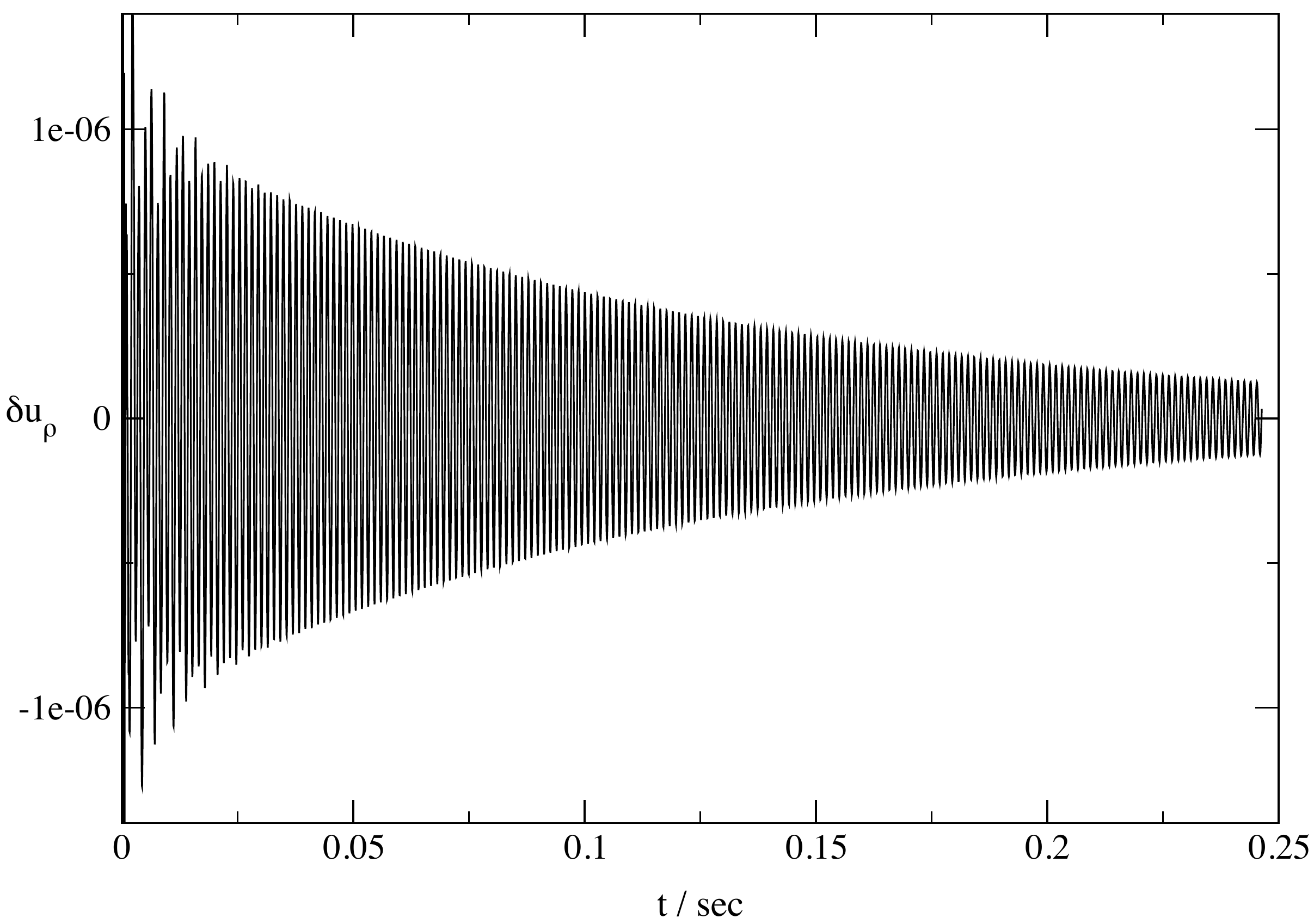}
\includegraphics[width=0.48\textwidth]{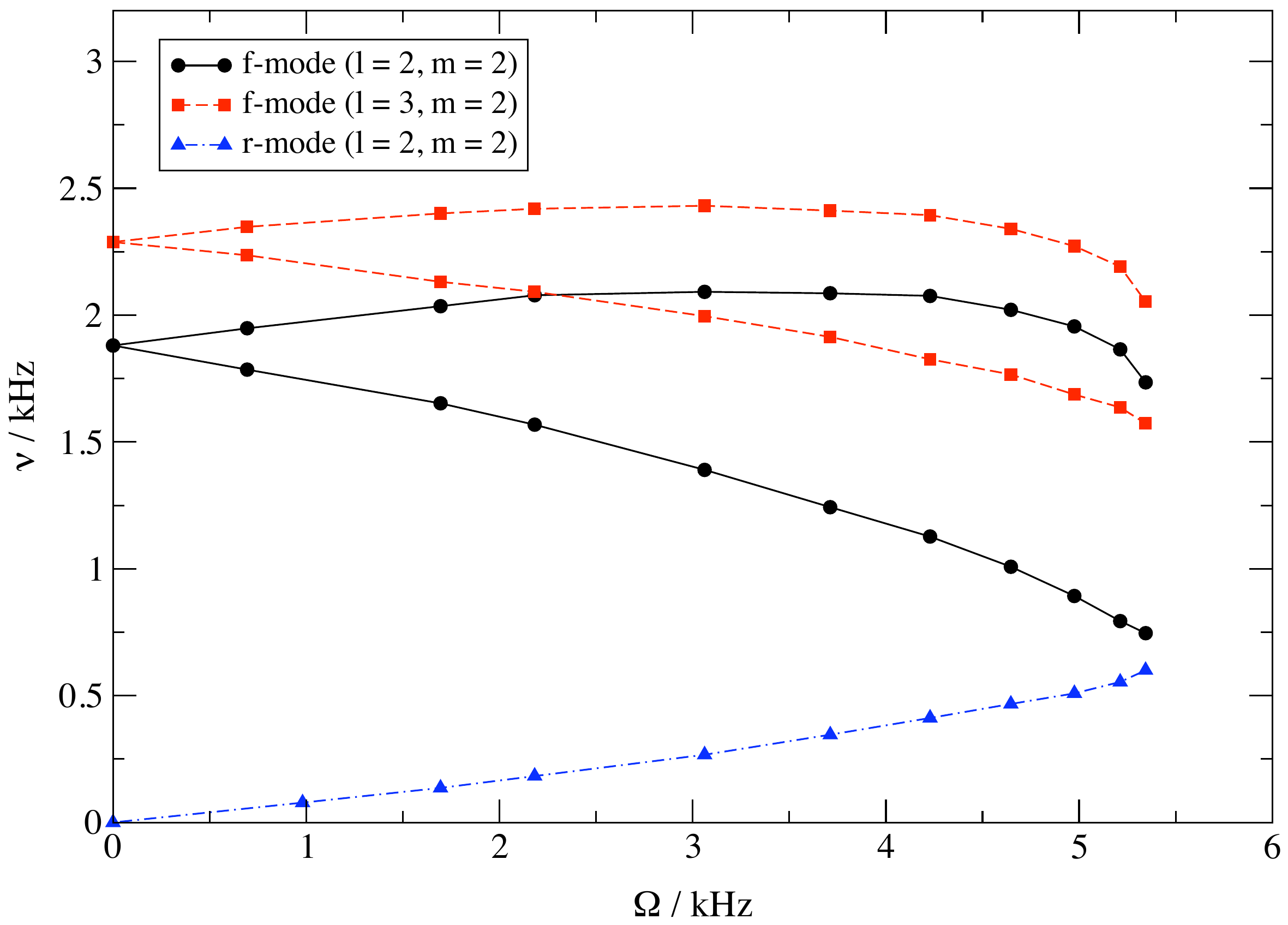}
\caption{{\it Left panel:} Time evolution of the perturbed $\varrho$-component of the 4-velocity using the fastest spinning background model BU8 running for about 250 milliseconds. {\it Right panel:} Mode splitting for two fundamental, nonaxisymmetric pressure modes as well as the change in frequency for the purely axial $l = m = 2$ inertial mode; all frequencies are shown in the comoving frame of reference.}
\label{fig:baroRun}
\end{figure}

A thorough test of the numerical implementation of system \eqref{eq:explicit_v2} in the barotropic case, i.e. $\Gamma = \Gamma_1$, can already be found in GK08 since the time evolution equations for stratified stars derived in this paper reduce to the form studied in GK08 for barotropic oscillations (see equation (11) there). Figure \ref{fig:baroRun} shows some typical results of such a simulation. The left panel depicts the time evolution of a perturbed velocity component for the most rapidly rotating stellar model BU8. Keep in mind, that we have to use artificial viscosity in order to stabilize the code against spurious modes. This also allows for very long integration times; in the case of Figure \ref{fig:baroRun} the simulation was cancelled after 0.25 seconds although longer runs pose no problems. The artificial viscosity has a secondary effect though. In addition to the numerical dissipation that is always present when using finite-difference representations of differential equations, the extra viscosity amplifies this dissipative effect which can clearly be seen in the decreasing amplitude of the initial perturbation. In frequency space, this typically leads to a broadening of the less dominant peaks in the power spectrum. However, subsequent mode recycling runs for the specific mode of interest can countervail this behaviour.

\subsection{Stratified Oscillations}
\label{ssec:stratifiedOscillations}
Once the perturbations of the fluid obey a different equation of state than the unperturbed background, buoyancy can drive a new class of oscillation modes. It acts in a similar manner as usual convection, i.e. perturbed fluid elements rise in the star as long as their densities are smaller than the unperturbed surroundings. As soon as the densities are equal, there is no net force acting on the fluid elements but due to their inertia, they overshoot this equilbrium position and move into regions where they are heavier than the surroundings. They begin to fall, overshoot again but now into regions where they are lighter and the cycle starts again.

\begin{figure}[ht!]
\centering
\includegraphics[width=0.46\textwidth]{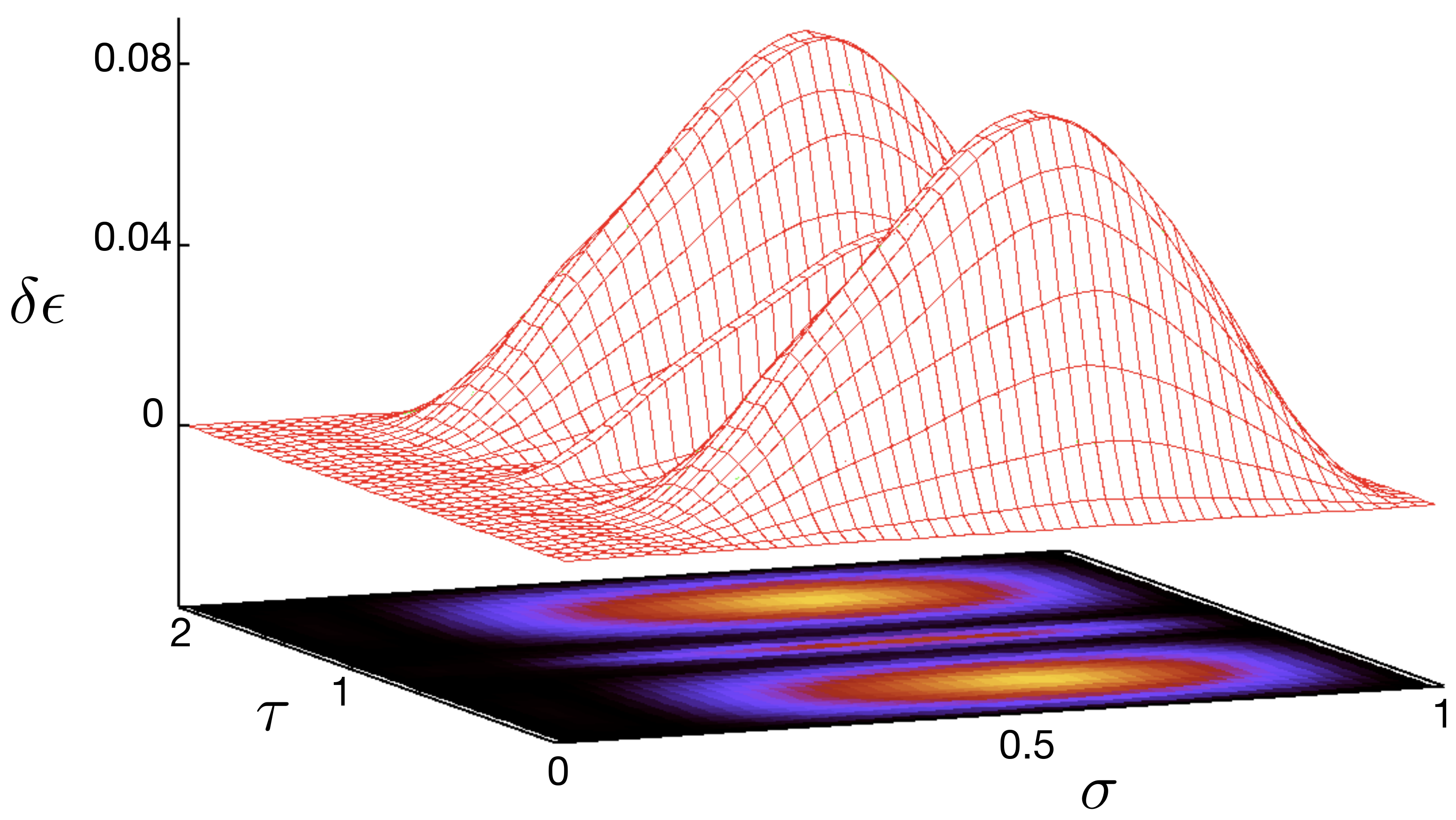}
\includegraphics[width=0.52\textwidth]{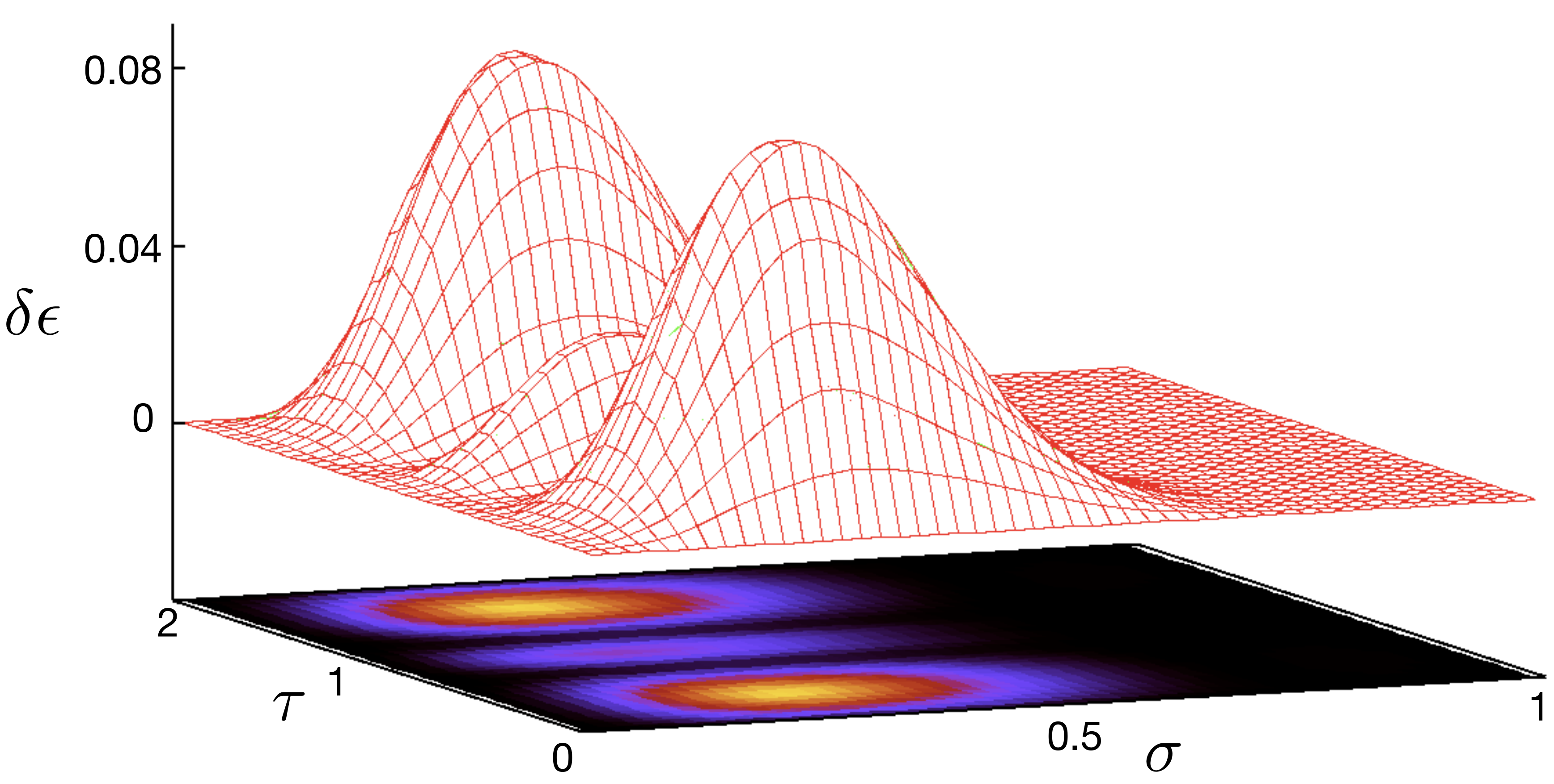}
\caption{Comparison between pressure- and gravity-mode eigenfunctions. {\it Left panel:} Energy-density eigenfunction for the $l=4, m=2$ fundamental  p-mode. {\it Right panel:} Energy-density eigenfunction for the $l=4, m=2$ fundamental g-mode.}
\label{fig:efComparison}
\end{figure}

It is already well known from Newtonian calculations (see e.g. \cite{Unno:1989kx}), that p- and g-modes differ significantly both in their frequencies and their eigenfunctions. While the former have frequencies at around 1.5 kHz and higher for typical equations of state and are virtually confined in a region between an inner reflection point and the surface of the star, g-modes have much lower frequencies, depending on the degree of stratification, and the large amplitudes that they reach in the stellar interior decrease pretty strong towards the neutron stars' surface. Figure \ref{fig:efComparison} shows a comparison between the two-dimensional eigenfunctions of the fundamental p- and g-mode, reconstructed from our relativistic time-evolution code and a composition gradient of $\Gamma_1 = 2.05$.

In the case depicted in Figure \ref{fig:efComparison}, the simulation was performed on the non-rotating BU0 model with trial initial data that excited the $^4{\mathrm f}$-mode at a frequency of 2.6 kHz as well as the $^4{\mathrm g}_1$-mode at $304$ Hz. The layout of the computational domain and the physical meaning of its coordinates $(\sigma, \tau)$ were already described in GK08. Essentially $\sigma$ is a radial coordinate ranging from the origin of the star to its surface while $\tau$ is an angular coordinate that reaches from one pole of the star, i.e. $\theta = 0$ in a spherical system, to the other one at $\theta = \pi$, where $\tau = 1$ corresponds to the equatorial plane. As one can see clearly, the maximum amplitude of the pressure mode is indeed located near the surface of the star while the g-mode eigenfunction has its peak value near the center of the star and falls off quickly towards the surface.

In order to further check the reliability of our code with respect to these new non-barotropic oscillations only found in stratified stars, we computed the frequencies of the fundamental $l = 2,3,4$ g-mode for a nonrotating, weakly stratified star with $\Gamma = 2.0$ and $\Gamma_1 = 2.0004$. A similar calculation has been performed in \cite{Yoshida:2000kx} within the Newtonian framework. The following Table \ref{tab:YLresults} summarizes the results. The overall agreement is very good, although it should be noted that the small value of $\Gamma_1$ leads to correspondingly low frequencies for the fundamental g-modes. Together with the constant rest-mass density of the BU model series of $\rho_{c} = 0.79\times 10^{15}$ g/cm$^3$, this means that the absolute values of the three g-mode frequencies lies within a bandwith of $\nu\approx 20 - 30 $ Hz. It is very demanding to accurately determine the frequency peaks in this small frequency range. High order modes even have smaller frequencies; according to \cite{Yoshida:2000kx}, the first overtones for $l = 2,3,4$ can be found in the range $\nu \approx 14 - 20$ Hz. Currently, the numerical noise in our code prevents us from properly determining these low frequencies.

\begin{table}
\centering
\caption{Comparison of the first $l = 2,3,4$ fundamental g-modes with the results from Yoshida \& Lee (2000). The frequencies are given in units of $\sigma/\sqrt{G\rho_c}$, where $\rho_c$ is the central rest-mass density. The literature values are denoted with YL.}
\label{tab:YLresults}
\begin{tabular}{ccc}
\hline
\;\;l\;\; 	& \;\;$g_1$\;\;	& \;\;$g_1$\;\;\\
		& YL			&		\\
\hline
2		& 0.0189		& 0.0181 \\
3		& 0.0227		& 0.0223 \\
4		& 0.0255		& 0.0254\\
\hline
\end{tabular}
\end{table}

In \cite{Passamonti:2009zr}, the fundamental $l = 2$ g-mode as well as several overtones were computed for different values of $\Gamma_1$. These values go as high as $\Gamma_1 = 2.4$ and lead to considerably higher frequencies for stratified oscillations. We calculated the mode frequencies for the fundamental and the first overtone of the $l = 2$ g-mode of the non-rotating and non-barotropic BU0 model for which $\Gamma = 2.0$ and checked it again with the literature values. The results are depicted in Table \ref{tab:PHAJHresults} and show very good agreement; only for larger composition gradients, i.e. larger $\Gamma_1$, there are some larger deviations. Actually, the authors in \cite{Passamonti:2009zr} used a relativistic, time-independent eigenvalue formulation for slowly rotating stars in the Cowling approximation developed and described in \cite{Passamonti:2008vn} and a new Newtonian time-evolution approach for rapidly rotating configurations to show the good agreement between these two independent methods in the case of a nonrotating star. The relative difference between these two schemes never exceeds $7\%$. The time evolutions performed with our code agree a little bit better with the eigenvalue results, which should be no surprise since for the special case of a nonrotating star, both relativistic codes are solving the same problem; either as boundary value problem as in \cite{Passamonti:2008vn} or as a time-dependent evolution problem presented in this paper.

\begin{table}
\centering
\caption{Comparison of the fundamental and first overtone $l = 2$ g-mode with the Passamonti et al. (2009) results. The frequencies are given in units of $\sigma/\sqrt{G\rho_c}$, where $\rho_c$ is the central rest-mass density. The literature values from the boundary value approach are denoted with PHAJH. Additionally, the absolute values of the mode frequencies (in Hz) for the model BU0 are shown in parentheses.}
\label{tab:PHAJHresults}
\begin{tabular}{ccccc}
\hline
\;\;$\Gamma_1$\;\;	& \;\;$^2g_1$\;\;		& \;\;$^2g_1$\;\;		& \;\;$^2g_2$\;\;		& \;\;$^2g_2$\;\;\\
				& PHAJH			&			& PHAJH		&		\\
\hline
2.05		&	0.195	&	0.196 (227)		&	0.134	&	0.132 (153)\\
2.2		&	0.383	&	0.382 (442)		&	0.266	&	0.258 (299)\\
2.4		&	0.531	&	0.503 (582)		&	0.370	&	0.344 (398)\\	
\hline
\end{tabular}
\end{table}

To summarize, the results compared so far for nonrotating stars and different values of the composition gradient are consistent with previous studies within the error bounds. We can now turn to the investigation of rotational effects on the frequencies of g-modes and study the influence of stratification on these frequencies.

From the nonrotating and slow-rotation limit it is already well known that the angular part of the fluid perturbations and, in the case of a coupled spacetime evolution, the corresponding metric perturbations, can be decomposed in terms of scalar, vector and tensor spherical harmonics which split into two distinct classes depending on their behaviour under space reflection, i.e. $(r, \theta, \varphi)\rightarrow (r, \pi - \theta, \pi + \varphi)$. Polar perturbations that are expressed in terms of $Y_{lm}$ and its gradient $\nabla Y_{lm}$ transform as $(-1)^l$ while axial perturbations behave like $r\times \nabla Y_{lm}$ and transform with $(-1)^{l+1}$. In a non-rotating, barotropic, perfect fluid star there are no non-trivial, i.e. zero-frequency, axial modes. In the case of the Cowling approximation, the only modes present are the pressure modes which have polar parity and frequencies of around 1 kHz and higher (see Figure \ref{fig:baroRun}). This changes, once rotation or a composition gradient is included. Both cases will introduce a new class of modes; inertial modes which are restored by the Coriolis force and the g-modes where gravity is the restoring force due to buoyancy.

In contrast to inertial modes which are degenerate for a nonrotating background star, gravity modes have a nonvanishing oscillation frequency already in the nonrotating limit. The following Figure \ref{fig:BaroVSStratEvolution} shows a comparison between fluid perturbations of a barotropic BU0-model and a stratified BU0 star with a composition gradient of $\Gamma_1 = 2.1$. The two upper panels show an 0.02 secs excerpt from a time-evolution performed for these different configurations.

\begin{figure}[ht!]
\centering
\includegraphics[width=0.48\textwidth]{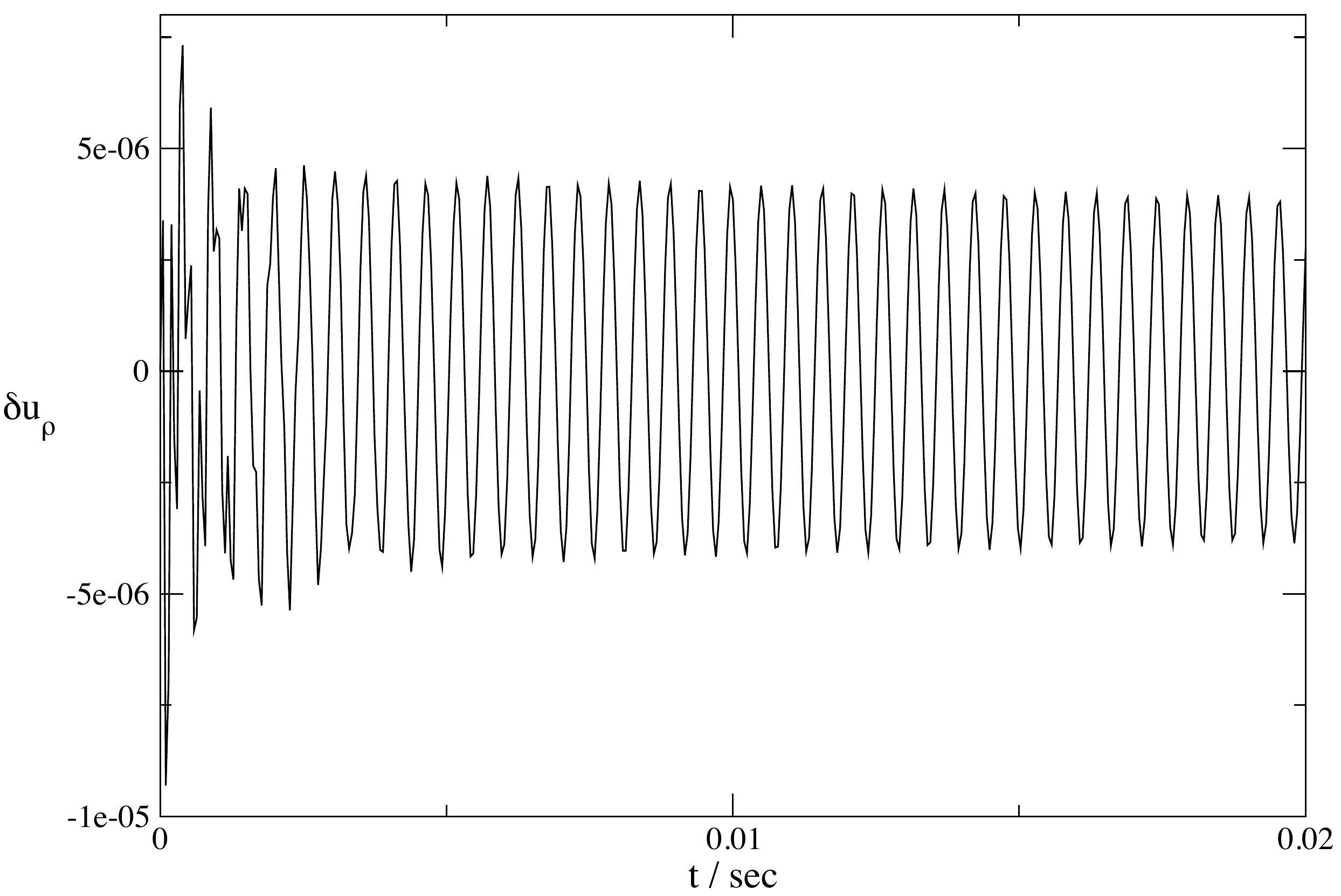}
\includegraphics[width=0.48\textwidth]{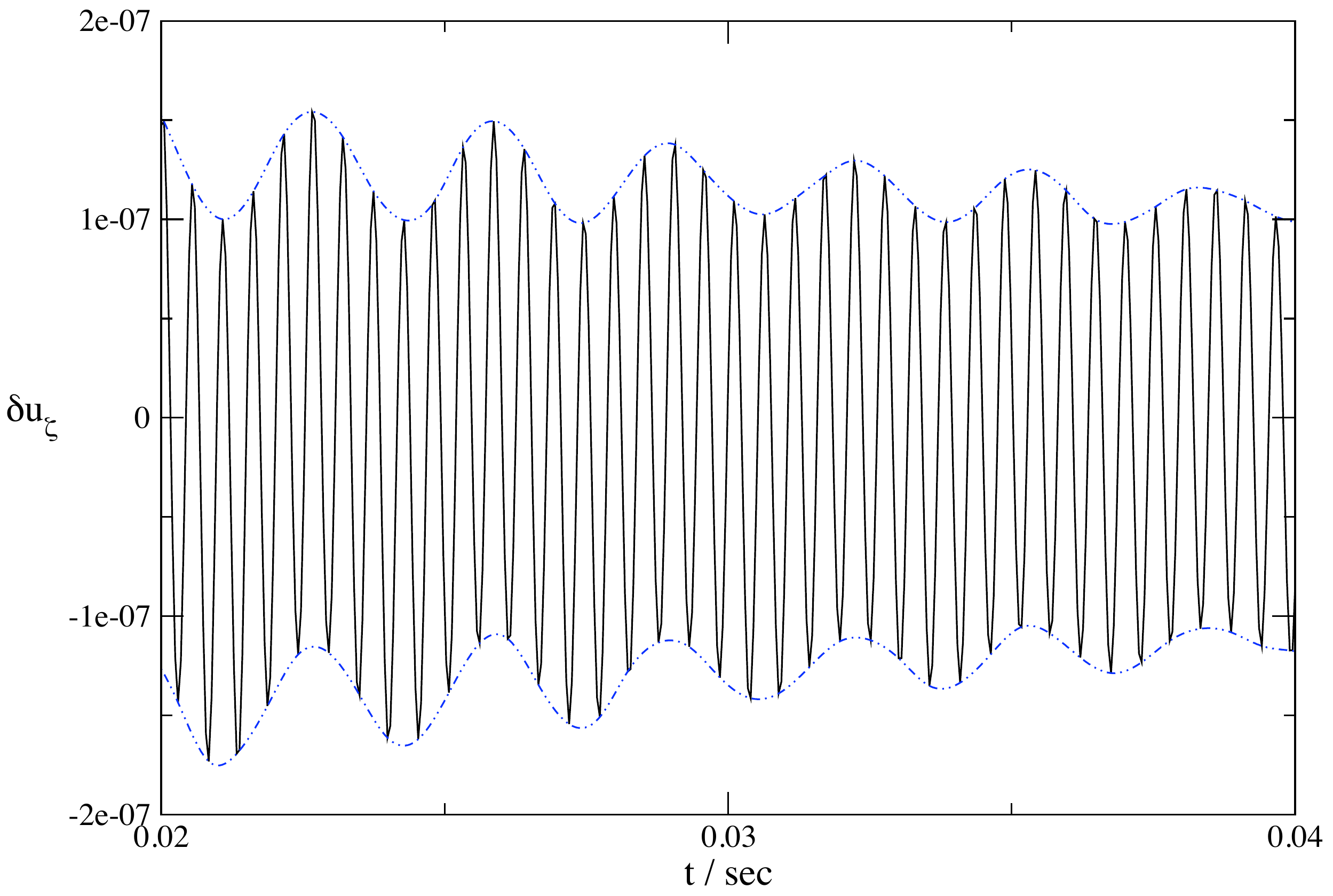}\\[0.1in]
\includegraphics[width=0.47\textwidth]{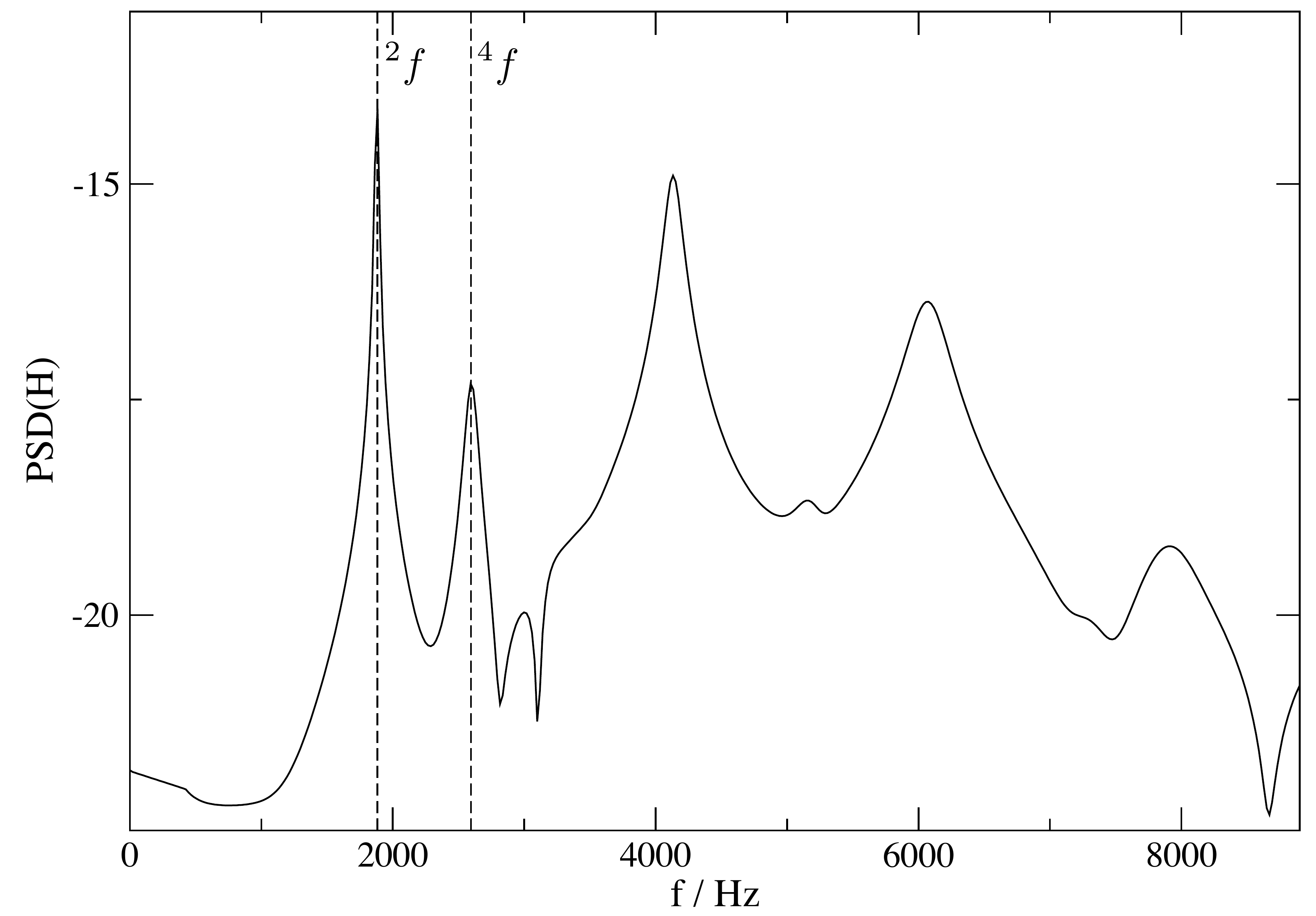}
\includegraphics[width=0.47\textwidth]{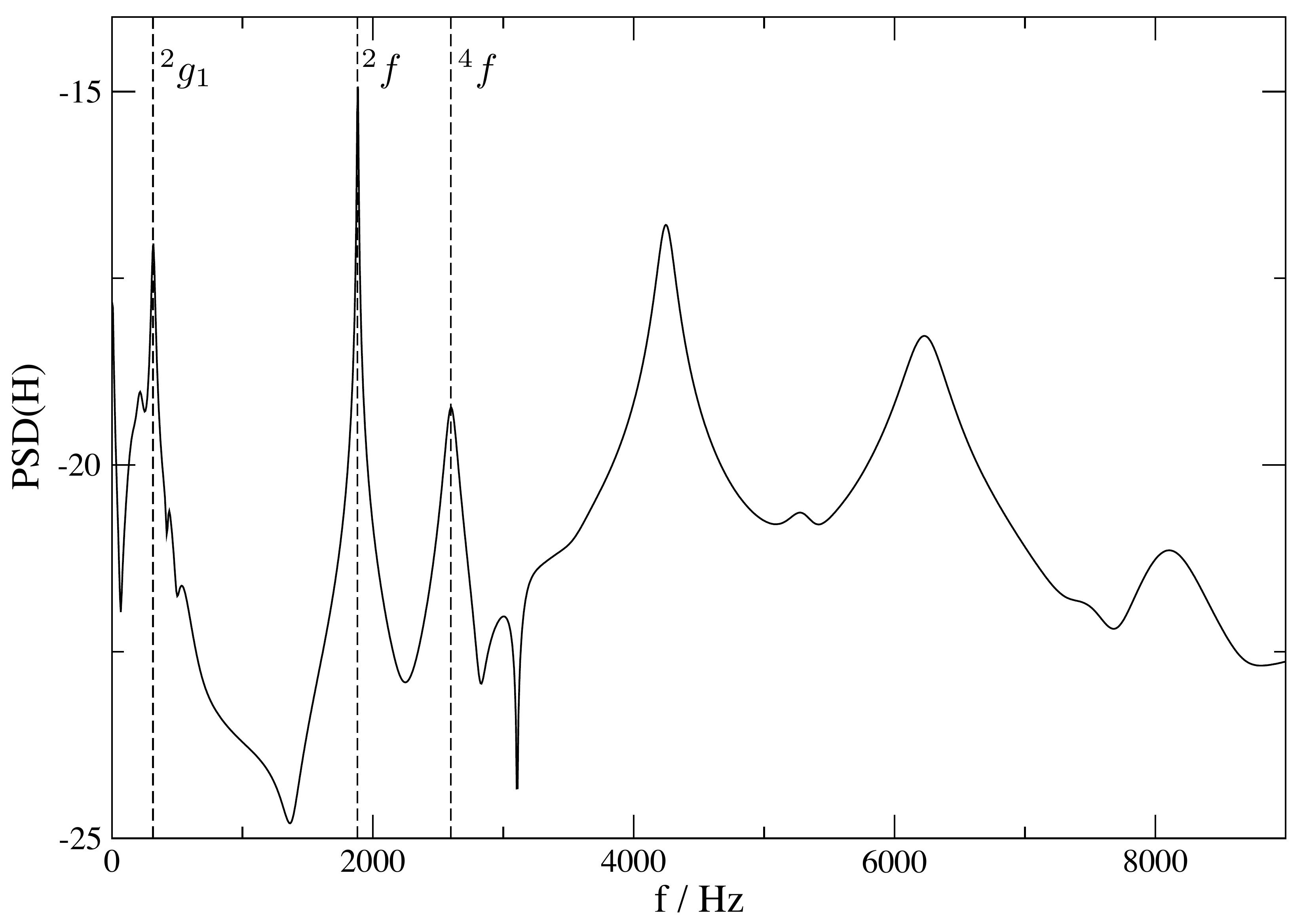}
\caption{Time evolution of fluid perturbations for a barotropic and a stratified equilibrium model in the nonrotating limit. {\it Top left panel:} Oscillation of $\delta u_{\varrho}$ for a barotropic neutron star with $\Gamma = \Gamma_1 = 2.0$. {\it Top right panel:} Oscillation of $\delta u_{\zeta}$ for a stratified neutron star with $\Gamma = 2.0$ and $\Gamma_1 = 2.1$; the modulation in the velocity perturbation is due to the influence of the fundamental $^2{\mathrm g}_1$-mode. {\it Bottom left panel:} The power spectral density for the barotropic model. {\it Bottom right panel:} The corresponding PSD plot for the stratified star.}
\label{fig:BaroVSStratEvolution}
\end{figure}

In both cases one can count roughly 37 oscillations of the dominant mode during this time which leads to an estimation of the corresponding frequency of about 1.85 kHz and in fact can be identified with the fundamental quadrupolar mode (see also Figure \ref{fig:baroRun}). But in the stratified case depicted in the right part of Figure \ref{fig:BaroVSStratEvolution}, there is an additional low-frequency modulation at around 300 Hz which is not present in the barotropic time-evolution and which is the fundamental quadrupolar g-mode at 314 Hz.

\begin{figure}[ht!]
\centering
\includegraphics[width=0.5\textwidth]{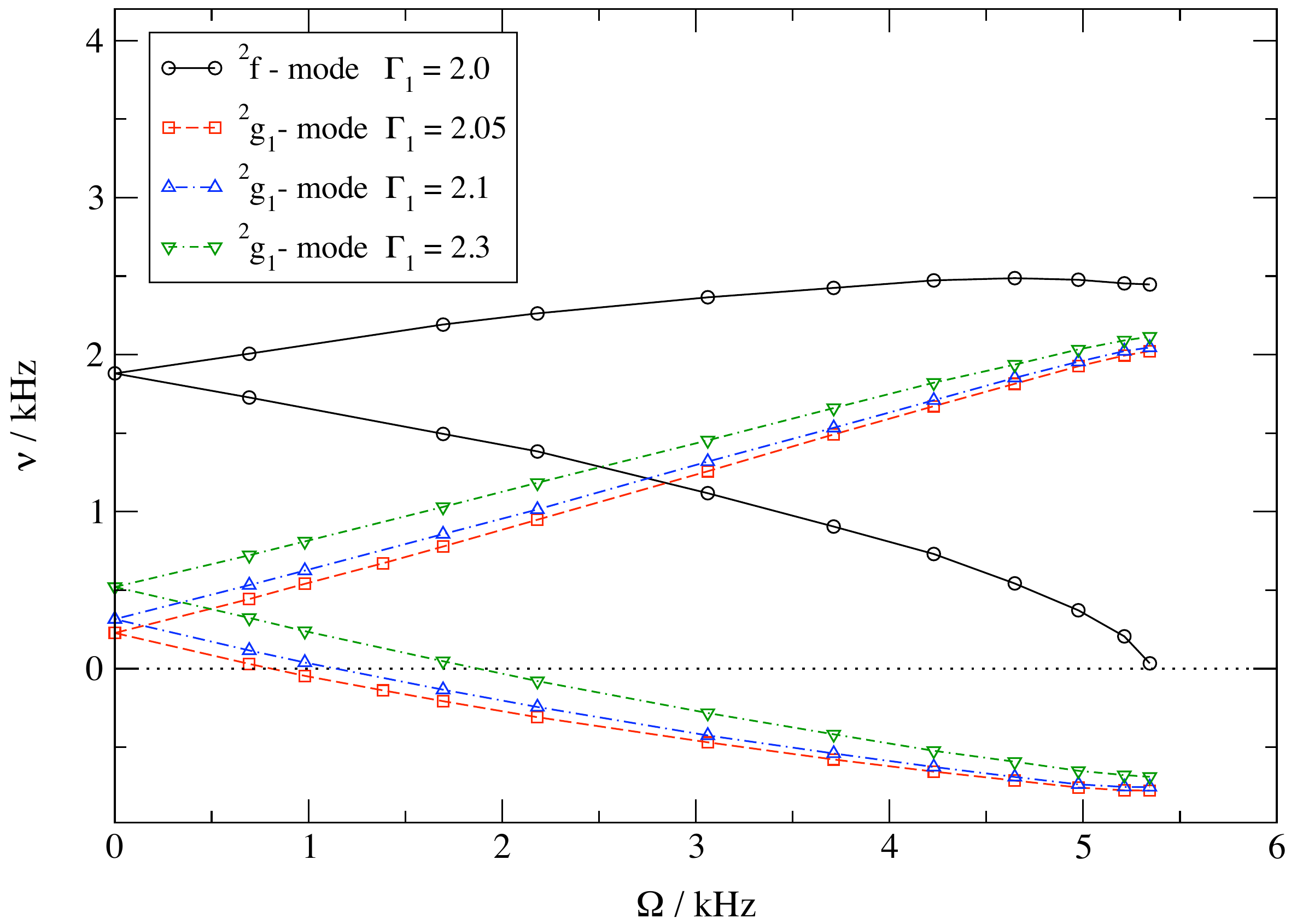}
\includegraphics[width=0.49\textwidth]{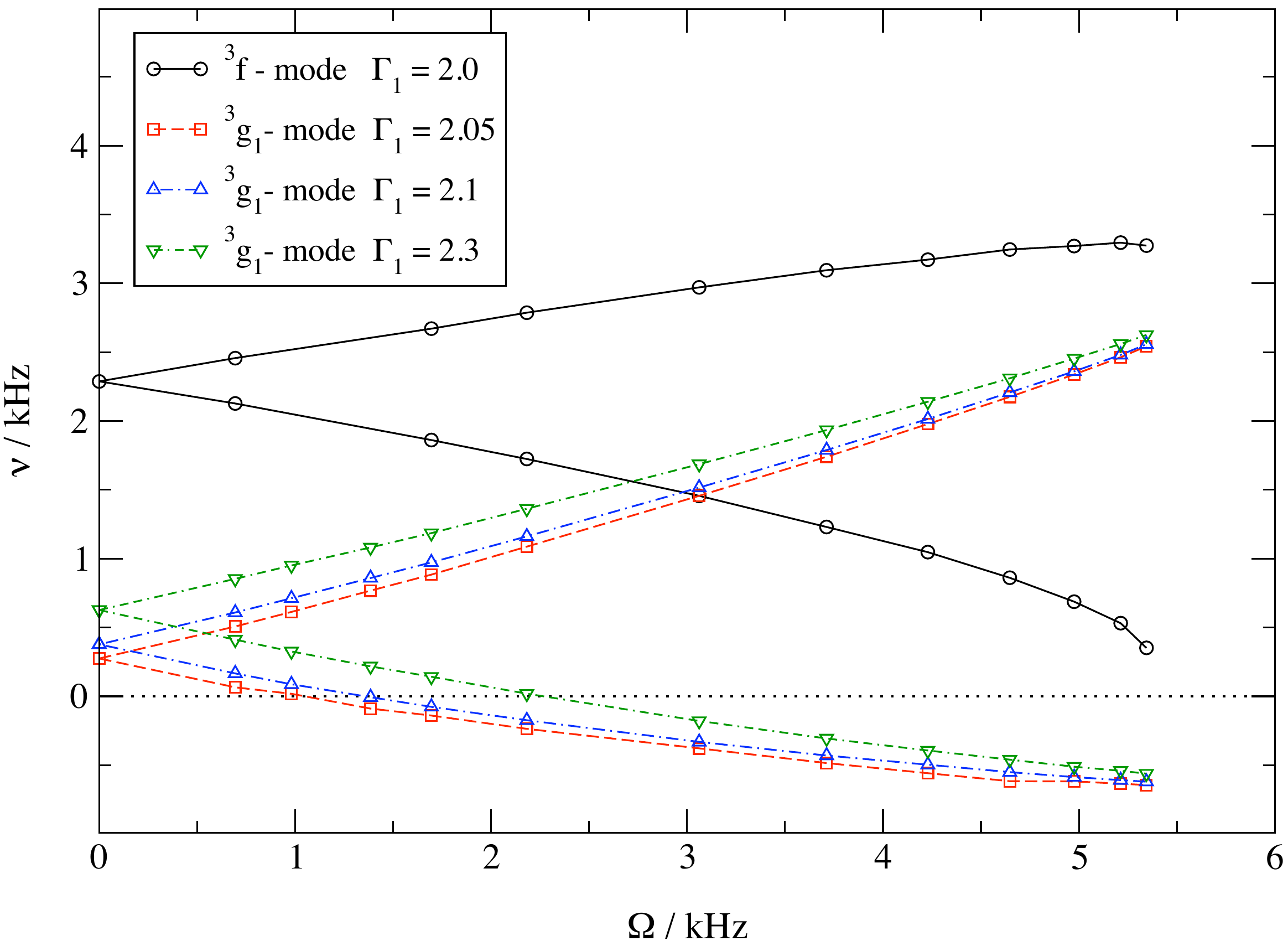}
\caption{Variation of mode frequencies for non-barotropic rotating stellar models with $\Gamma = 2.0$ and several values of $\Gamma_1$. {\it Left panel:} Frequencies for the $l=m=2$ barotropic f-mode and fundamental g-mode for different composition gradients. {\it Right panel:} Frequencies for the $l=3, m=2$ barotropic f-mode and fundamental g-mode for different composition gradients; all frequencies are shown in the inertial frame of reference.}
\label{fig:gModesWithRotation}
\end{figure}

Both mode classes, gravity- and inertial modes, have mixed polar and axial components and both of them occupy the low-frequency regime (see Figure \ref{fig:baroRun} for the purely axial $l=m=2$ inertial mode and Table \ref{tab:YLresults}, \ref{tab:PHAJHresults} for g-modes). Figure \ref{fig:gModesWithRotation} shows the variation of the fundamental g-mode frequency for several values of the composition gradient $\Gamma_1$. The left panel shows the polar-led $^2{\mathrm g}_1$-mode while the right panel depicts the axial-led $^3{\mathrm g}_1$-mode. Both panels show the oscillation frequencies in the inertial frame and for comparison, the barotropic fundamental pressure mode is also displayed for the corresponding harmonic indices. In both cases, the g-modes behave in a similar fashion. The higher the composition gradient $\Gamma_1$, the higher are also the mode frequencies; this can clearly be seen from Figure \ref{fig:gModesWithRotation}.

In contrast, the pressure modes only show a negligible effect on the magnitude of stratification, although their frequencies are also slightly larger. For example, in the nonrotating case, the barotropic $^2{\mathrm f}$-mode has a frequency of $\nu = 1.879$ kHz. For $\Gamma_1 = 2.05$ this changes to $\nu = 1.88$ kHz and is absolutely neglectable while for $\Gamma_1 = 2.3$ the frequency shifts to $\nu = 1.886$ kHz which is still only a relative shift of 0.4\%. This picture gets modified though a little bit in the case of rotation, mostly because the counterrotating mode moves down in frequency and therefore the relative difference increases. However, this difference never exceeds the 7\%-level. The following Figure \ref{fig:compareBaroVsStratFMode} shows a comparison between the frequency variation of the fundamental quadrupolar $^2{\mathrm f}$-mode with rotation for a purely barotropic star and a stratified background model with a comparatively large composition gradient of $\Gamma_1 = 2.3$. As already stated, the effect of stratification on pressure modes are minute.

\begin{figure}[ht!]
\centering
\includegraphics[width=0.5\textwidth]{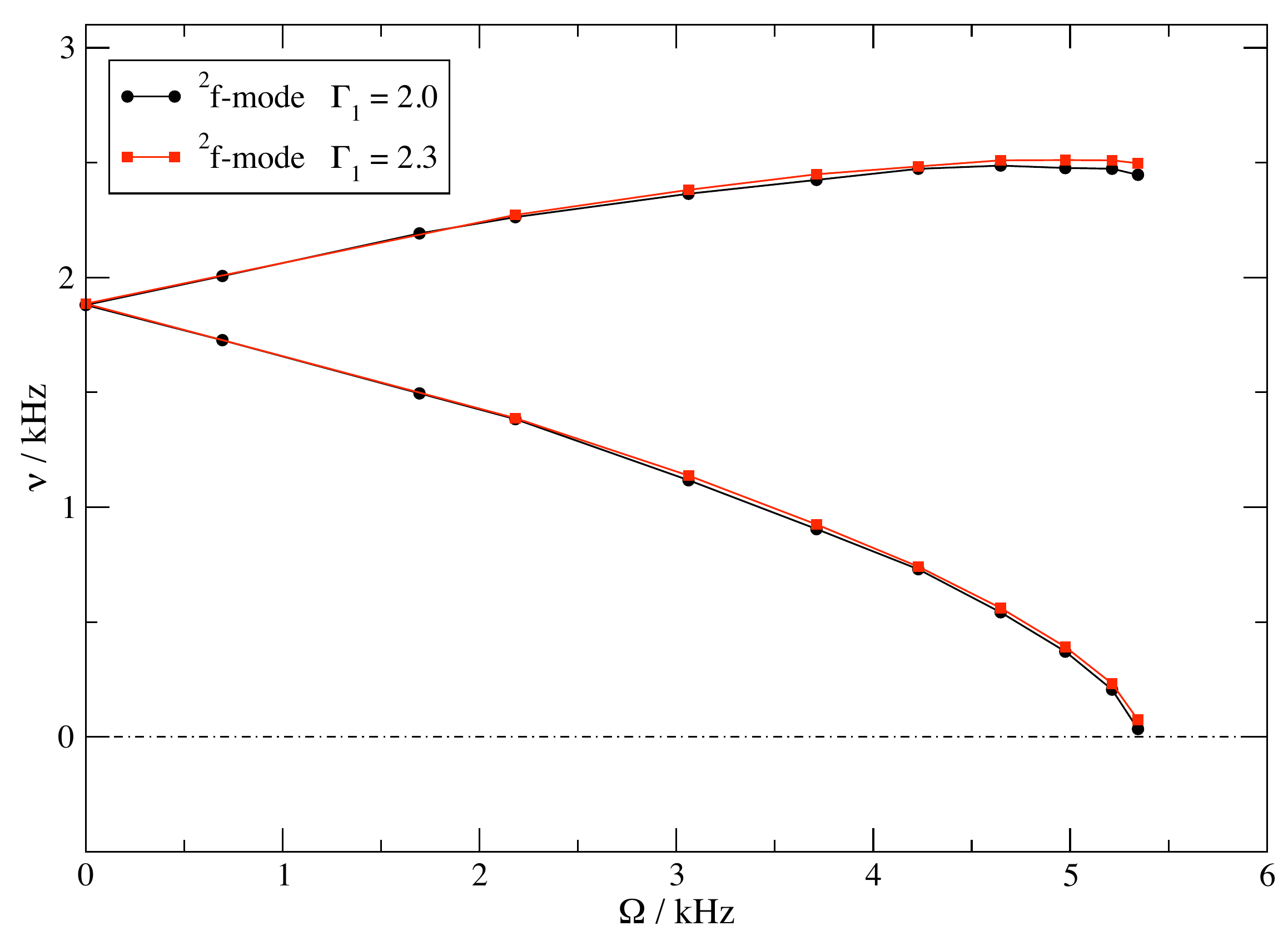}
\caption{The effect of stratification on the fundamental quadrupolar $^2{\mathrm f}$-mode. A rather huge composition gradient of $\Gamma_1 = 2.3$ only leads to a negligible shift of the mode frequencies to larger values.}
\label{fig:compareBaroVsStratFMode}
\end{figure}

As soon as the frequency of a specific mode drops down to zero, i.e. when it becomes degenerate, this oscillation mode is prone to the so-called CFS-instability. This type of rotational dragging instability was originally discovered in  \cite{Chandrasekhar:1970rt}, thoroughly investigated in \cite{Friedman:1978fr,Friedman:1978ys} and always works when there is a coupling mechanism to some radiation field (for example gravitational radiation) and a rotating background that leads to a splitting between co- and counterrotating modes. A rotational instability can only occur when the mode frequency vanishes along a sequence of equilibrium models.

It has already been observed in GK08, that the $^2{\mathrm f}$-mode of the model BU series goes marginally unstable just at the mass-shedding limit. This can also be seen from the left panel of Figure \ref{fig:gModesWithRotation} and Figure \ref{fig:compareBaroVsStratFMode}, where the dotted horizontal line represents the zero-frequency limit. The quadrupolar fundamental mode is a very good gravitational wave emitter, however it does not enter the instability unless the background star spins extremely fast. The absolute values for the non-axisymmetric $^2{\mathrm f}$-mode will change in a fully relativistic treatment of the problem with coupled spacetime evolution (see for example \cite{Dimmelmeier:2006zr}), but qualitatively the picture remains unchanged; the CFS-unstable quadrupolar f-mode usually only works in rapidly rotating stars, if at all.

On the other hand, due to their low frequency, the fundamental g-mode becomes unstable at low rotation rates as can be clearly seen in Figure \ref{fig:gModesWithRotation}. Depending on the degree of stratification, the critical spin frequency increases for higher composition gradients since the mode frequencies are also growing in this case. In \cite{Lai:1999yq}, Lai gives a spin frequency estimation for the onset of the secular g-mode instability depending on the nonrotating oscillation frequency $\sigma_0$ which is
\begin{equation}
\label{eq:laiEstimate}
\Omega_c = 0.68\,\sigma_0\,,
\end{equation}
where $\sigma_0 = 2\pi\nu_0$. For the six unstable modes in Figure  \ref{fig:gModesWithRotation} we did least square fittings with all available data points and a quadratic fitting polynomial. We find a somewhat smaller value than 0.68 in \cite{Lai:1999yq} which is $\Omega_c = c\,\sigma_0$ with $c = 0.58\pm 0.01$. This is roughly 17\% smaller than the original value and agrees very well with Newtonian results which lead to a similar decrease. However, it should also be pointed out that the g-mode instability is not thought of being a very significant gravitational wave emitter; viscosity will damp away the instability very soon.
Actually, the growth times for unstable g-modes, in the slow rotation approximation and by using a post-Newtonian approach, have already been calculated in \cite{Ferrari:2003qu}. The results there suggest growth times of the order of 10$^5$ - 10$^9$ secs, which is considerably longer than the time viscosity will need to suppress  completely the instability.

For stratified stars, there is now the interesting situation, that two classes of modes practically occupy the same domain in frequency space. Both emerge from degenerate, zero frequency oscillations and as the rotation rate of the background star increases, both classes will be affected by the growing Coriolis force. Inertial modes are inherently dominated by the Coriolis force but of course this force will also influence the g-modes. For these type of oscillation, buoyancy is the restoring force but the faster the compact object is spinning, the more important the Coriolis force will become. How dominant it gets essentially depends on two parameters. First of all there is the rotation rate; the Coriolis force will grow with increasing angular velocity of the background star. One would therefore expect that in the case of rapidly spinning background models, the g-modes will become more similar to pure barotropic inertial modes and this in fact verified by Newtonian calculations; see again \cite{Passamonti:2009zr}. However, the rotation rate and hence the strength of the Coriolis force is limited by the mass-shedding limit; it cannot grow beyond the breakup frequency. This is where the second parameter, which determines the interplay beween g-modes and inertial modes, becomes important; it is the degree of stratification. In the case discussed in this paper, where polytropic equations of state are used for both the static equilibrium configuration and the perturbations, it is the difference between $\Gamma_1$ and $\Gamma$ that is crucial. The larger this difference is the larger is also the density difference between unperturbed and perturbed fluid and therefore the larger is the effect of buoyancy. This means that if we just choose the polytropic index of the perturbed fluid $\Gamma_1$ high enough, then the influence of the increasing Coriolis force will be counterbalanced and possibly overtaken by the buoyant force. In this case, the similarities between g-modes and barotropic inertial modes concerning both frequencies and eigenfunctions may not be so pronounced as it is the case for smaller composition gradients.

\begin{figure}
\centering
\includegraphics[width=0.6\textwidth]{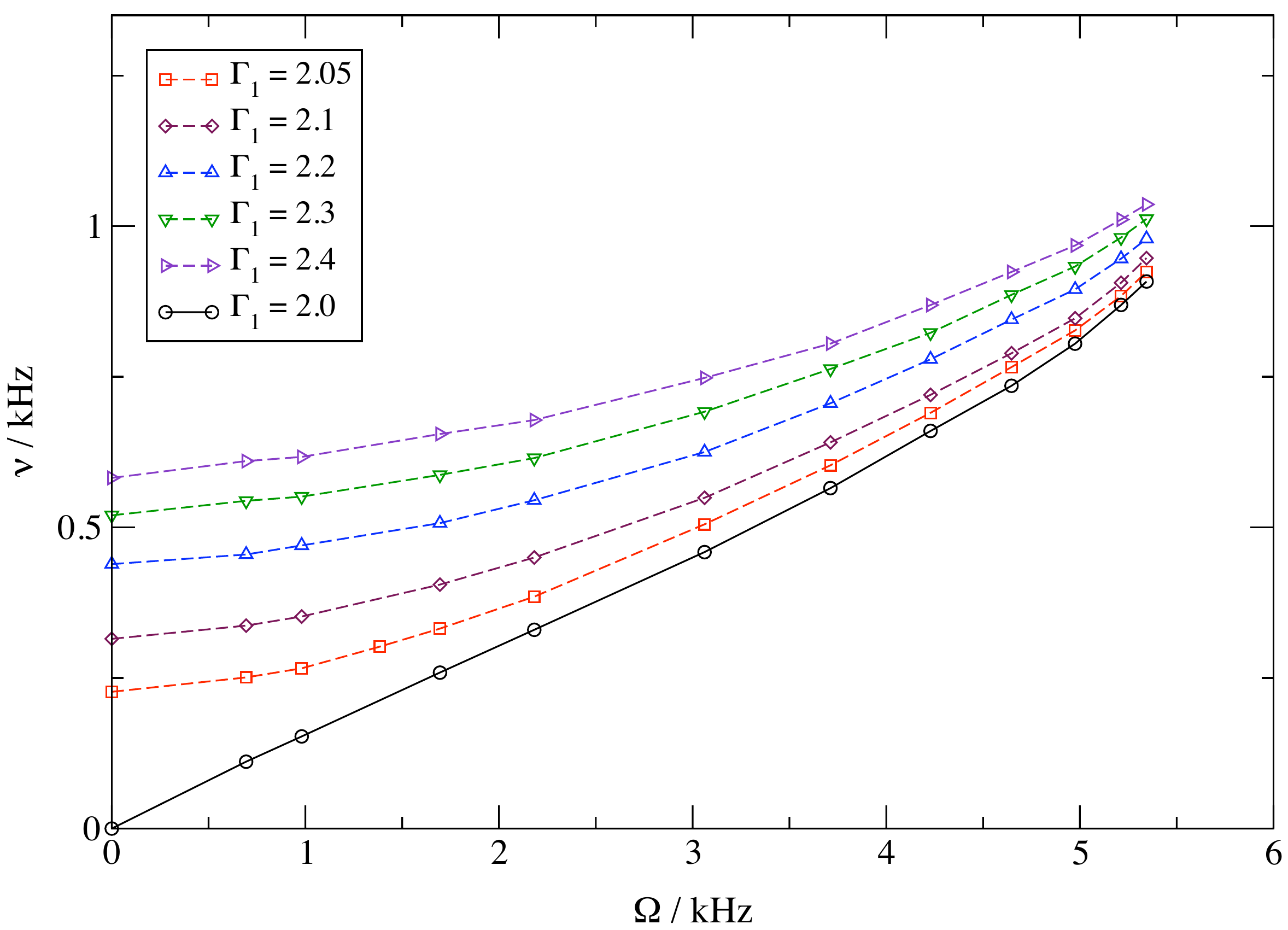}
\caption{Variation of mode frequencies with the rotation rate of the background star for the counterrotating $^2g_1$-mode and several composition gradients $\Gamma_1$. Additionally, a purely barotropic inertial mode for which $\Gamma_1 = \Gamma$ is also depicted and shows how g- and inertial mode frequencies approach for high rotation rates and small composition gradients; all frequencies are shown in a comoving frame of reference.}
\label{fig:baroAndGmodes}
\end{figure}

We checked this expected behaviour with the unstable, counterrotating fundamental $^2{\mathrm g}_1$-mode and various degrees of stratification. Since in this case the mode is travelling in opposite direction with respect to the rotation of the star, it easily gets dragged forward at a rather low spin frequency and is therefore affected by the CFS-instability. In the left panel of Figure \ref{fig:gModesWithRotation}, these unstable g-modes are depicted in an inertial frame of reference and we are focusing now on the unstable branches at low frequencies. Figure \ref{fig:baroAndGmodes} shows the result of the simulations for composition gradients ranging from $\Gamma_1 = 2.0$ for the barotropic inertial mode to $\Gamma_1 =  2.4$ for the gravity mode with the highest degree of stratification. In contrast to Figure \ref{fig:gModesWithRotation}, all frequencies are depicted in a reference frame corotating with the star. Figure \ref{fig:baroAndGmodes} essentially confirms our assumptions about the interplay between g- and inertial modes for growing rotation rates and increasing Coriolis force opposed to the growing influence of the buoyant force for larger composition gradients. As already discussed earlier, the g-mode frequencies increase for larger degrees of stratification and they also increase for higher rotation rates. For small composition gradients, i.e. with $\Gamma_1 = 2.05$ and $\Gamma_1 =  2.1$ in Figure  \ref{fig:baroAndGmodes}, the mass-shedding limit for our chosen background models is high enough for the Coriolis force to dominate over buoyancy in the rapid rotation regime. Consequently, the g-modes become similar to certain barotropic inertial modes which in particular means that their oscillation frequencies approach each other. The Coriolis force is also relevant for g-modes with a higher degree of stratification but in these cases, i.e. with $\Gamma_1$ roughly starting from a value of 2.2 in Figure \ref{fig:baroAndGmodes}, the effect of buoyancy is strong enough even at the Kepler-limit to counteract the Coriolis force. This leads to a somwhat larger deviation of the g-mode frequencies from the purely barotropic inertial mode frequencies as can be clearly seen in Figure \ref{fig:baroAndGmodes}.

The similarity between inertial and g-modes for a sufficiently large Coriolis force does not only affect the oscillation frequencies but also the mode eigenfunctions. In fact, this is how we are able to find the barotropic inertial mode that closestly approaches the $\Gamma_1 = 2.05$ counterrotating $^2{\mathrm g}_1$-mode. For this, one typically starts at a rather high rotation rate of the background model and tries to excite the fundamental g-mode. Several mode-recycling runs are performed to obtain an accurate frequency determination. Then the eigenfunctions for the various perturbation variables are extracted and put back as initial data but now for barotropic oscillations with $\Gamma_1 = \Gamma$. This will usually excite some inertial modes within the expected frequency range which then again have to be identified by their eigenfunction. Once the correct oscillation mode has been picked out, one can follow its frequency variation towards lower and higher rotation rates.

\begin{figure}
\centering
\includegraphics[width=0.95\textwidth]{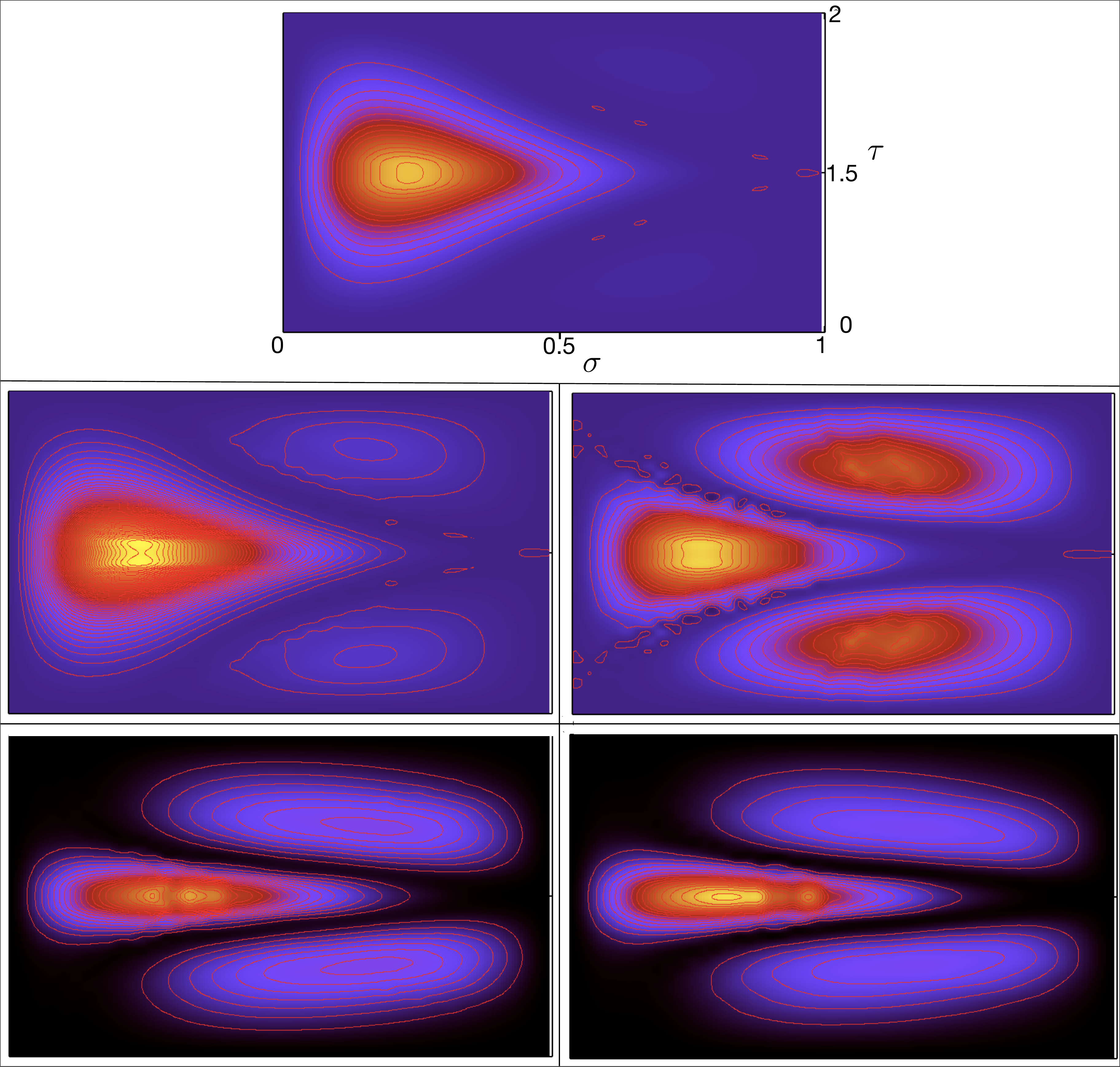}
\caption{Eigenfunctions of the $\delta u_\varrho$-perturbation for the counterrotating $^2g_1$-mode and the barotropic inertial mode at several rotation rates. {\it First row:} The eigenfunction of the gravity mode for model BU0. {\it Second row, left panel:} The eigenfunction of the fundamental gravity mode for model BU01. {\it Second row, right panel:} The corresponding eigenfunction of the barotropic inertial mode for model BU01. {\it Last row, left panel:} The eigenfunction of the fundamental gravity mode for model BU6. {\it Last row, right panel:} The eigenfunction of the inertial mode for model BU6.} 
\label{fig:eigenfunctionComparison}
\end{figure}

Figure \ref{fig:eigenfunctionComparison} shows the results of such simulations for certain eigenfunctions at several rotation rates. The depiction is similar to the eigenfunction plots from Figure \ref{fig:efComparison}, but this time the amplitude of the eigenfunction is projected onto the two-dimensional computational domain rather than additionally plotted along a third axis. Instead, contour lines are used to underline the characteristics of the various eigenfunctions. The coordinates on the numerical domain are identical to the ones used in Figure \ref{fig:efComparison}, i.e. $\sigma$ represents the radial coordinate and ranges from 0 to 1 while $\tau$ is the the angular coordinate and ranges from 0 to 2. The topmost row shows the $\delta u_\varrho$-eigenfunction of the fundamental counterrotating $^2{\mathrm g}_1$-mode for $\Gamma_1 = 2.05$ and a nonrotating background model at a frequency of $\nu_g= 227$ Hz. Since there are no inertial modes for $\Omega = 0$, there is also no corresponding eigenfunction. The two panels in the second row show again the eigenfunction for $\delta u_\varrho$ but this time for model BU01 rotating at $\Omega = 694$ Hz. The left picture depicts once more the fundamental quadrupolar g-mode now at $\nu_g = 251$ Hz where one can already see the formation of two smaller secondary peaks located symmetrically above and below the equatorial plane. The right panel in the same row shows the corresponding eigenfunction of the barotropic inertial mode at the same rotation rate and with a frequency of $\nu_i = 111$ Hz. Here, the secondary peaks are much more pronounced and almost equal in amplitude compared to the main peak. One can already figure out a certain global resemblance between the inertial and g-mode; at least the major characteristics are similar in both cases. This similarity becomes even more evident at very high rotation rates. The last two panels in the bottom row depict the very same eigenfunctions but now at a spin frequency of $\Omega = 4.976$ kHz; this is roughly 93\% of the mass-shedding limit. In accordance with the frequencies that approach each other for high rotation rates, see Figure \ref{fig:baroAndGmodes}, the same happens with the eigenfunctions. In the last two panels, the $^2{\mathrm g}_1$-mode has a frequency of $\nu_g = 827$ Hz while the barotropic inertial mode oscillates at $\nu_i = 805$ Hz. In both cases, the main peak gets narrowed in a confined region along the equatorial plane while the secondary peaks grow in amplitude for the g-mode and decrease for the inertial mode. By this gradual transition, the eigenfunctions resemble each other very good at high spin frequencies.

\begin{figure}
\centering
\includegraphics[width=0.6\textwidth]{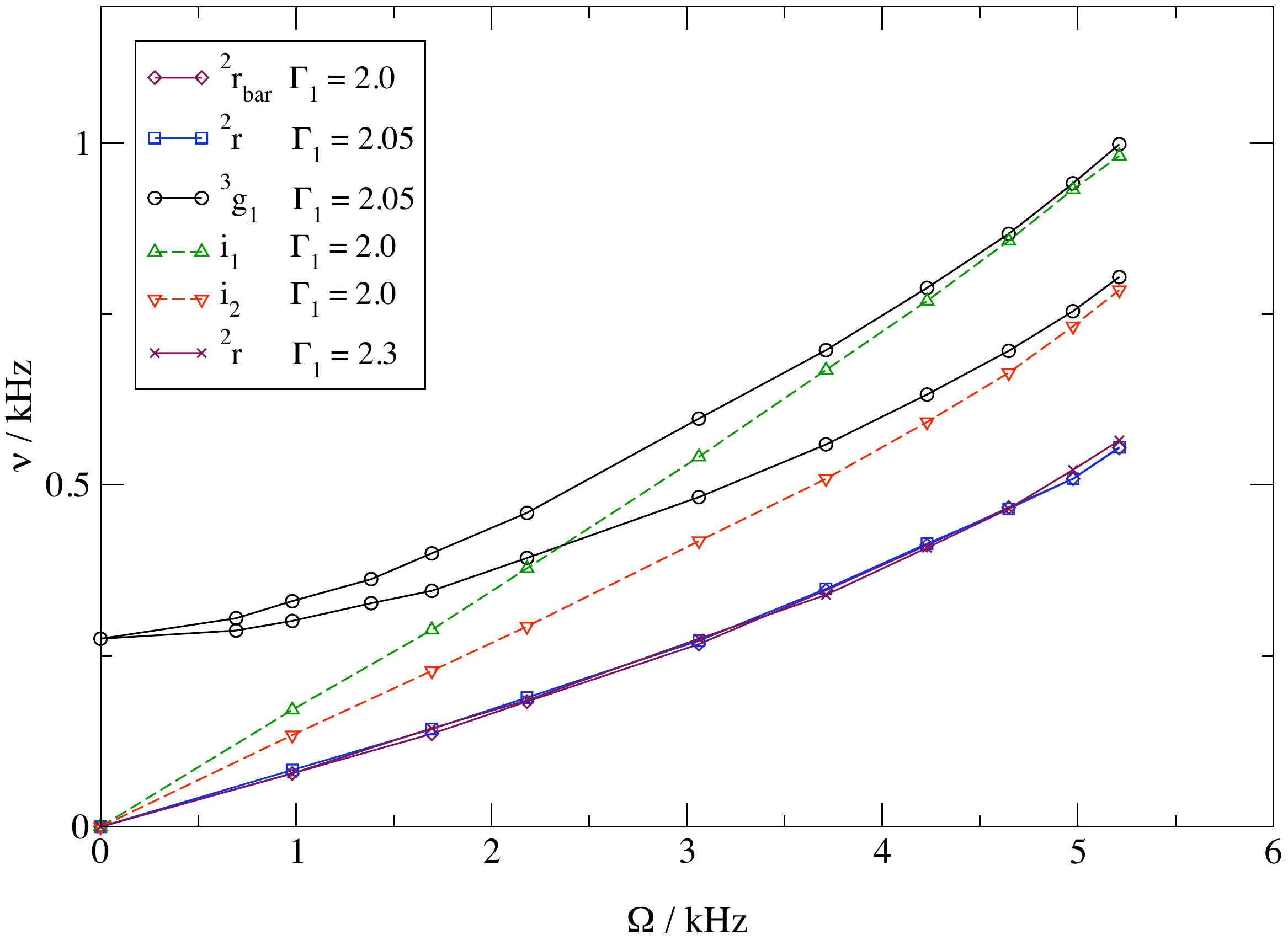}
\caption{The frequencies of the co- and counterrotating $^3{\mathrm g}_1$-mode as well as the barotropic inertial modes ${\mathrm i}_1,{\mathrm i}_2$ which approach the gravity modes for high rotation rates as seen from a comoving coordinate system. Similar to polar-led g-modes, the increasing Coriolis force dominates over buoyancy for small composition gradients. In contrast, the purely axial r-mode shown here is essentially unaffected by the degree of stratification.} 
\label{fig:axialLedModes}
\end{figure}

A similar observation can be made for axial-led g-modes like the $^3{\mathrm g}_1$-mode, whose frequency variation can be seen in the right panel of Figure \ref{fig:gModesWithRotation}. Figure \ref{fig:axialLedModes} shows both the counter- and corotating branches of this mode for $\Gamma_1 = 2.05$ but now in a coordinate system comoving with the star. As already described in GK08, the high frequency branch can be identified with the retrograde rotating oscillations, i.e. these modes become unstable once the rotation rates exceeds a certain threshold, while the low frequency branch corresponds to modes that travel in the prograde direction. What is also depicted in Figure \ref{fig:axialLedModes} are the two barotropic inertial modes that approach both of these branches as the spin frequency increases. As for polar-led g-modes, the rather small degree of stratification leads to a dominance of the Coriolis force over buoyancy at sufficiently large rotation rates and therefore the g-modes approach eigenfrequencies of certain barotropic inertial modes. In addition, we also show in this picture the sensitivity of the purely axial $l=2, m=2$ r-mode frequencies on several composition gradients. As already observed in \cite{Passamonti:2009zr}, there is virtually no effect of the degree of stratification on the r-mode. Even for the rather large value of $\Gamma_1 = 2.3$ the change in frequency is essentially negligible.

\section{Conclusions}
\label{sec:conclusions}

In this work it as been presented, for the first time, a study of the g-modes for fast rotating stars in the general relativistic framework. The results are in qualitative agreement with those by \cite{Passamonti:2009zr} which were derived also for fast rotating stars but in the Newtonian framework.

This study demonstrates that potentially we are able to analyze future observational data from neutron stars in the general relativistic framework for equilibrium models which where constructed by using Einstein's theory of gravity. 

One of the main results observed in this work is the mixing of the g-modes with inertial modes for high rotation rates which suggests that it will be difficult for asteroseismology to discriminate the effects of stratification. On the other hand, for lower rotation rates the degree of stratification may become observable since the two families of modes maintain their distinct character. 

As a next step we plan to study the effect of differential rotation on all families of fluid modes. It is expected that differential rotation will be present during the very first moments of the formation of neutron stars when their fluid actually will be stratified. In this case the neutron stars are still quite warm, i.e. temperatures considerably higher than  $10^9$K while the matter is in the form of degenerate protons $p$, neutrons $n$ and electrons $e$ which are not yet superfluid. This will be an extension of this work and the work that has been done for differentially and slowly rotating stars in \cite{Passamonti:2008vn}.

\section{Acknowledgments}
\label{sec:acknowledgements}
We thank Hajime Sotani for helpful discussions. This work was also supported
by the German Science Foundation (DFG) via SFB/TR7.

\bibliographystyle{apsrev}
\bibliography{references}

\end{document}